\documentclass{mn2e}
\usepackage{epsfig}
\usepackage{lscape}
\usepackage{rotating}
\usepackage{longtable}


\def\ls{{_<\atop^{\sim}}}
\def\gs{{_>\atop^{\sim}}}

\begin{document}

\title{X-Ray Detection of Warm Ionized Matter in the Galactic Halo}

\author[F. Nicastro et al.]{F. Nicastro,$^{1,2,3}$ F. Senatore,$^{1,4}$ A. Gupta,$^{5,6}$ M. Guainazzi,$^{7}$ S. Mathur,$^{5,8}$ 
\newauthor 
Y. Krongold,$^9$ M. Elvis,$^2$ L. Piro,$^{10}$ \\
$^1$Osservatorio Astronomico di Roma - INAF, Via di Frascati 33, 00040, Monte Porzio Catone, RM, Italy \\
$^2$Harvard-Smithsonian Center for Astrophysics, 60 Garden St., MS-04, Cambridge, MA 02138, USA \\
$^3$University of Crete, Heraklion, Greece \\
$^4$ Universit\'a degli Studi di Roma ``Tor Vergata'', Via Orazio Raimondo, 18 - 00173 Roma, Italy \\
$^5$Ohio State University, Columbus, OH, USA \\
$^6$Columbus State Community college, Columbus, OH, USA \\
$^7$ European Space Astronomy Centre of ESA, Villanueva de Caada, Madrid, Spain \\
$^8$Center for Cosmology and Astro-Particle Physics (CCAPP), The Ohio State University, OH, USA \\
$^9$Instituto de Astronomia, Universidad Nacional Autonoma de Mexico, Mexico City, Mexico \\
$^{10}$Istituto di Astrofsica e Planetologia Spaziali - INAF, Roma, Italy}

\maketitle
\begin{abstract}
We report on a systematic investigation of the cold and mildly ionized gaseous baryonic metal components of our Galaxy, through 
the analysis of high resolution {\em Chandra} and XMM-{\em Newton} spectra of two samples of Galactic and extragalactic sources. 
The comparison between lines of sight towards sources located in the disk of our Galaxy and extragalactcic sources, allows us for the first time to 
clearly distinguish between gaseous metal components in the disk and halo of our Galaxy. 
We find that a Warm Ionized Metal Medium (WIMM) permeates a large volume above and below the Galaxy's disk, perhaps up to the Circum-Galactic 
space (CGM). This halo-WIMM imprints virtually the totality of the OI and OII absorption seen in the spectra of our extragalactic targets, has a temperature 
of T$_{WIMM}^{Halo}=2900 \pm 900$ K, a density $<n_H>_{WIMM}^{Halo} = 0.023 \pm 0.009$ cm$^{-3}$ and a metallicity $Z_{WIMM}^{Halo} = (0.4 \pm 0.1)$ $Z_{\odot}$. 
Consistently with previous works, we also confirm that the disk of the Galaxy contains at least two distinct gaseous metal components, one 
cold and neutral (the CNMM: Cold Neutral Metal Medium) and one warm and mildly ionized, with the same temperature of the halo-WIMM, but higher 
density ($<n_H>_{WIMM}^{Disk} = 0.09 \pm 0.03$ cm$^{-3}$) and metallicity ($Z_{WIMM}^{Disk} = 0.8 \pm 0.1$ $Z_{\odot}$). 
By adopting a simple disk+sphere geometry for the Galaxy, we estimates masses of the CNMM and the total (disk + halo) WIMM of $M_{CNMM} \ls 8 \times 
10^8$ M$_{\odot}$ and $M_{WIMM} \simeq 8.2 \times 10^9$ M$_{\odot}$. 
\end{abstract}
\begin{keywords}
Absorption Lines, ISM, Galaxy, WHIM
\end{keywords}

\section{Introduction}
Astrophysical experiments aimed to X-ray the inter-stellar (ISM) or circum-galactic (CGM) medium of our 
own Galaxy along the lines of sight to Galactic and extragalactic X-ray sources, are not only powerful in diagnosing 
the physical, chemical and dynamical state of these, likely multi-phase and metal enriched, mediums but also offer the 
best avenue to benchmark position and strength of the thousands of electronic transitions from the shells K (N=1), L (N=2) and M (N=3) of abundant 
metals, from C to Fe (e.g. Verner, Verner \& Ferland, 1996; Garcia et al., 2005; Smith et al., 2001) that fill the soft X-ray band. 

The 6.2-124 \AA\ (0.1-2 keV) band contains indeed thousands of electronic transitions whose wavelengths and oscillator strengths, however, 
are accurately known only for a few tens, mostly those from the outermost occupied shells (e.g. K transitions from H-like 
and He-like ions of C to Fe; e.g. Verner, Verner \& Ferland, 1996). 
The so called inner-shell transitions from the same elements (e.g. K transitions of O ions from OI to OVI) are 
much more difficult to predict theoretically and only a handful of laboratory measurements exist (e.g. Savin et al., 
2012). Moreover all these calculations as well as the existing laboratory measurements, suffer large uncertainties both 
on wavelengths and strengths (e.g. Gatuzz et al., 2013a,b). 
Knowing the exact position and strength of these many resonant transitions is also vital, for example, to properly disentangle in the X-ray spectra 
of extragalactic targets, Galactic foreground absorption from putative intervening absorption systems and so avoid mis-identifications. 

Unfortunately, however, these experiments are tough, and mostly limited by the intrinsic weakness of the 
transitions coupled with the limited effective area and spectral resolution of the current X-ray spectrometers (the 
High- and Low-Energy-Transmission Gratings onboard {\em Chandra} - HETG, Canizares, Schattenburg \& Smith, 1986 and LETG, Brinkman et al., 1986 - 
and the Reflection Grating Spectrometer onboard XMM-{\em Newton} - RGS, Hettrick \& Kahn, 1986), but also by 
the presence of instrumental features (particularly in the RGS). 
For unresolved absorption line detectability, the spectrometer resolution and effective area play 
a complementary role, being present in the function that defines the minimum detectable line equivalent width (EW hereinafter) 
as the root square of their ratio: $EW_{thresh} = (N_{\sigma} \Delta \lambda_{RE}) / (\sqrt{F_{\lambda_0}^{ph} A_{eff}^{\lambda_0} T_{Exp} 
\Delta \lambda_{RE}})$, where $N_\sigma$ is the statistical significance, in mumber of standard deviations $\sigma$, down to 
which the threshold is evaluated, $\Delta \lambda_{RE}$ is the spectrometer Resolution Element in wavelength, $F_{\lambda_0}^{ph}$ is the 
source specific photon flux at $\lambda_0$, $A_{eff}^{\lambda_0}$ is the spectrometer effective area at $\lambda_0$ and $T_{Exp}$ is the 
exposure time of the observation. 
In the important O K transition region ($\lambda \simeq 21-24$ \AA) the HETG and LETG, have first-order 
effective areas of only 0.5-5 cm$^{2}$ and 8-15 cm$^2$, respectively, and Line Spread Function (LSF) 
full-width-half-maximum (FWHM)  of 310-370 km s$^{-1}$ and 620-740 km s$^{-1}$, respectively. 
The RGS has a larger effective area ($\sim 40$ cm$^{2}$ in the 21-24 \AA\ region, where only one of the two 
RGSs is operating), but a lower spectral resolution, with a highly non-gaussian LSF with FWHM$\simeq 
870-1000$ km s$^{-1}$ (e.g. Williams et al., 2006).  
Therefore very bright background sources or very long integration times (or both) are needed to detect these weak absorption lines fron the 
Galaxy ISM/CGM. 

\noindent
Only few of such experiments have been therefore successfully performed so far. In 2004, Juett and collaborators, analyzed 
the HETG spectra of seven Galactic binaries with S/N per resolution element (SNRE, hereinafter) of about 5 in the continuum 
in the region of the O K edge, and were able to identify several O K features from neutral (OI) and 
ionized (OII and OIII) ISM. 
In particular they determined a position of $\lambda = 23.508 \pm 0.003$ \AA\ for the OI K$\alpha$ transition, 
which implied a shift of about 50 m\AA\ compared to previous theoretical estimates by Gorczyca \& McLaughlin (2000) 
and Pradhan et al. (2003). They also identified the strong line ubiquitously present at $\lambda \simeq 23.35$ \AA\ as 
due, at least partly, to the OII K$\alpha$ transition from the mildly ionized ISM, contrary to what had been proposed 
previously by other authors (Paerels et al. 2001; Schulz et al. 2002; Takei et al. 2002), who identified this feature as 
due to iron dioxide. In 2009, Yao and collaborators confirmed the OI K$\alpha$ and OII K$\alpha$ identifications, 
at $\lambda = 23.51$ \AA\ and $\lambda = 23.35$ \AA\ respectively, in the higher quality spectrum (SNRE $\simeq 10-14$ 
in the O K region) of Cygnus X-2. 
Finally, recently Gatuzz and collaborators (2013a,b) studied the neutral and ionized ISM in the disk of our Galaxy 
in the high quality (SNRE$\sim 10-20$) HETG spectra of the low mass X-ray binary XTE J1817-330, and reported the detection of several K 
resonant lines from OI, OII and OIII In particular, for the first time they identified the weak K$\beta$ transition from OII in the Galaxy ISM, at 
$\lambda = 22.29$ \AA\ (Gatuzz et al., 2013b). 
In a subsequent work (Gatuzz et al., 2014), the same authors confirmed the presence of a mildly ionized (OI to OIII) ISM in the 
Galactic disk, along the lines of sight towards a sample of 8 Galactic binaries observed with the {\em Chandra} gratings. 
A similar work was presented by Pinto and collaborators (2013), who studied the Galactic disk ISM in both its gasous, molecular 
and dust components in the RGS spectra of 9 Galactic sources. 


\noindent
In this work for the first time we study the cold and mildly ionized ISM/CGM of our Galaxy by analyizing two distinct samples of Galactic and 
extragalactic sources. In particular we reduce and analyze RGS data from a S/N-limited (SNRE$\ge 10$ at 22.2 \AA) sample of 20 Galactic X-ray binaries 
and LETG/HETG data of the 28 extragalactic sources from the (non S/N-limited) parent sample of Gupta et  al. (2012), plus the RGS 
data of the blazar Mkn~501, and present a systematic study of the cold metal gaseous components of the Galaxy's disk and halo. 

The paper is organized as follows. In \S 2 we present our data reduction and analysis and discuss the results of our analysis for our Galactic 
and extragalactcic samples, separetely. 
Section 3 is instead dedicated to the modeling, discussion and interpretation of our results: we first derive the basic properties of the OII absorbers 
(for which we have 2 transitions) through a detailed curve of growth analysis of their K$\alpha$ and K$\beta$ transitions (\S 31.-3.5), and then dedicate 
\S3.6-3.7 to a self-consistent photoionization modelization of the Galactic and extragalactic lines of sight absorbers that includes both OI and OII. 
Finally, \S 4 summarizes our conclusions. 

\section{Data Reduction and Analysis}
We reduced and analyzed archival RGS data from a well-defined (see \S 2.1) sample of Galactic sources and LETG/HETG data of 
a number of extragalactic sources (the two blazars with the highest SNRE spectra available and, for comparison, 
the lower SNRE parent sample of Gupta et al., 2012: see \S 2.2) in the spectral region of the OI K edge (21-24 \AA). 
All data 
were reduced with the latest versions of the XMM-{\em Newton} and {\em Chandra} data reduction and 
analysis softwares (``Science Analysis System'', Gabriel et al., 2004 - SAS - v. 13.5.0, and 
``Chandra Interactive Analysis of Observation'', Fruscione \& Siemiginowska, 1999 - CIAO - v. 4.6.1) and calibrations 
(automatically set according to the given observation, for XMM-{\em Newton} data, and CALDB v. 4.6.2 for {\em Chandra}). 
We followed the appropriate XMM-{\em Newton} and {\em Chandra} data reduction/analysis threads and documentation to 
extract HETG/LETG and RGS spectra and responses for all observations and, when multiple observations existed 
for a single source, we co-added their spectra to maximize the SNRE. 

\subsection{The Galactic Sample}
Our ``Galactic'' sample is extracted from a total of 856 RGS observations of 131 low-mass and high-mass X-Ray binaries 
in our own Galaxy. After extracting the RGS spectra of all 131 sources from these 
856 observations, we selected only the Galactic targets with SNRE$\ge 10$ at 22.2 \AA\ in their co-added RGS spectrum.  
This reduced the initial sample of 131 sources to 20 objects. 
Table 1 lists the 20 targets (ordered by increasing 3$\sigma$ equivalent width detectability threshold - i.e. decreasing sensitivity - for 
unresolved absorption lines: EW$_{thresh} = 3 \times \Delta\lambda /$SNRE), together with: (a) their Galactic line-of-sight column density of 
neutral H as inferred from 21-cm measurements (Kalberla et al., 2005; second column) and (b) as measured by us through spectral fitting of 
the RGS spectra (third column), (c) their Galactic latitude $b$ (fourth column), (d) their distance (when available; fifth column), 
(e) the spectra SNREs (sixth column) and (f) the 3$\sigma$ EW$_{thresh}$ at 22.2 \AA\ (seventh column). 
\begin{table*} 
\begin{minipage}{126mm}
\caption{\bf Galactic Sources}
\begin{tabular}{|ccccccc|}
\hline
Source Name & N$_H$ (21 cm) & N$_H^X$ & Galactic Latitude $b$ & Distance & SNRE$^a$ & 3$\sigma$ EW$_{Thrresh}$ \\
& in 10$^{20}$ cm$^{-2}$ & in 10$^{20}$ cm$^{-2}$ & in degree & in kpc & at 22.2 \AA\ & at 22.2 \AA, in m\AA\ \\ 
\hline
Her~X-1 & 2 & $2 \pm 1$ & +37.523 & $^b$6.6 & 65 & 3.2 \\
EXO~0748-676 & 10 & $6 \pm 1$ & -19.811 & $^c$8.0; $^d$5.7 & 58 & 3.6 \\
PSRB~0833-45 & 81 & $3 \pm 1$ & -2.787 & $^e$0.3 & 54 & 3.9 \\
SAX~J1808.4-3658 & 11 & $12 \pm 1$ & -8.148 & $^d$2.8 & 52 & 4.0 \\
Swift~J1753.5-0127 & 17 & $23 \pm 1$ & +12.186 & --- & 45 & 4.7 \\
Cygnus~X-2 & 19 & $20 \pm 1$ & -11.316 & $^c$13.4; $^d$11.0 & 43 & 4.9 \\
MAXI~J0556-332 & 3 & $4 \pm 2$ & -25.183 & ---  & 41 & 5.1 \\
Cygnus~X-1 & 72 & $53 \pm 1$ & +3.067 & $^f$2.1 & 37 & 5.7 \\
SWIFT~J1910.2-0546 & 23 & $35 \pm 1$ & -6.844 & --- & 38 & 5.5 \\
4U~1636-53 & 28 & $37 \pm 1$ & -4.818 & $^d$6.0 & 31 & 7.7 \\
4U~1728-16 & 20 & $21 \pm 1$ & +9.038 & $^g$4.4; $^h$7.5 & 26 & 8.1 \\
V*V821Ara & 37 & $42 \pm 1$ & -4.326 & $^ i$10.0 & 25 & 8.4 \\
GS~1826-238 & 17 & $33 \pm 1$ & -6.088 & $^j$7.5 & 20 & 10.5 \\
HETE~J1900.1-2455 & 10 & $13 \pm 2$ & -12.873 & $^k$5.0; $^d$3.6 & 21 & 10.0 \\
4U~2129+12 & 6 & $8 \pm 4$ & -27.312 & $^d$5.8 & 19 & 11.1 \\
4U~1543-624 & 24 & $22 \pm 2$ & -6.337 & $^l$7.0 & 18 & 11.7 \\
Aql~X-1 & 28 & $39 \pm 1$ & -4.143 & $^c$5.2; $^d$3.9 & 17 & 12.4 \\
4U~1735-444 & 26 & $29 \pm 1$ & -6.994 & $^c$9.4; $^d$6.5 & 16 & 13.1 \\
X~Persei & 7 & $20 \pm 2$ & -17.136 & $^n$0.8 & 12 & 17.5 \\
XTE~J1650-500 & 43 & $61 \pm 1$ & -3.427 & --- & 13 & 16.2 \\
\hline
\end{tabular}
$^a$ We assume 70 m\AA\ for the width of the RGS resolution element. 
$^b$ Reynolds et al. (1997; MNRAS \, 288, 43, REF.). 
$^c$ Jonkers, P.G. \& Nelemans, (2004; G., MNRAS, 354, 355, REF.). 
$^d$ Galloway, D.K. et al. (2008; ApJS, 179, 360, REF.)
$^e$ Caraveo, P.A., De Luca, A., Mignani, R.P. \& Bignami, G.F. (2001; ApJ, 561, 930, REF.). 
$^f$ Ziolkowski, L. (2001' MNRAS, 358, 851, REF.). 
$^g$ Grimm, H.J., Gilfanov, M. \& Sunyaev, R. (2002; A\& A, 391, 923, REF.). 
$^h$ Pinto, C., Kaastra, J.S., Costantini, E. \& de Vries, C. (2013, A\& A, 551, 25, REF.)
$^i$ Kawai, M. \& Suzuki, M. (2005, ATel, 534, 1, REF.). 
$^j$ Kong, A.K.H. et al., (2000, MNRAS, 311, 405, REF.).
$^k$ Hynes, R.I. et al. (2004, ApJ, 609, 317, REF.). 
$^l$ Wang, Z. \& Chakrabarty, D., (2004, ApJ, 616, L139, REF.). 
$^m$ Naik, S. \& Paul, B. (2004, A\& A, 418, 655, REF.). 
$^n$ Megier A.; Strobel A.; Galazutdinov G.A.; Krelowski J. (2009, A\& A, 507, 833, REF.)
\end{minipage}
\end{table*} 

\subsection{The Extragalactic Targets}
Due to their intrinsically lower soft X-ray flux, compared to that of not-too-highly absorbed Galactic X-ray binaries, generally 
extragalactic sources have archival grating spectra with SNRE$<< 10$. This makes it difficult to build up sizeable 
samples of extragalactic grating spectra with SNRE similar to that of the 20 sources of our Galactic sample. 
 
However, there are important exceptions: the total ACIS-S-LETG and HRC-S-LETG spectra of the two {\em Chandra} 
(and XMM-{\em Newton}) calibration (because they are blazars and so expected to have intrinsically featureless 
spectra) targets Mkn~421 ($z=0.03$) and PKS~2155-304 ($z=0.116$), are the highest quality high-resolution X-ray 
spectra ever taken,  with SNRE=80 and 74 for the ACIS-S-LETG and the HRC-S-LETG spectra of Mkn ~421, and SNRE=60 
and 49 for the ACIS-S-LETG and HRC-S-LETG spectra of PKS~2155-304, respectively. 
We therefore use these two sources (with their four spectra) as the reference targets of our extragalactic sample. 
In addition, to make the size of our extragalactic sample comparable to that of our Galactic sample, we include the 
26 sources of the Gupta et al. (2012: hereinafter G-sample) sample with LETG or HETG spectra with SNRE in the range 1.2-18.7, at 22.2 \AA\
\footnote{The total number of sources with ``good SNRE'' ($\ge 1$) at 22 \AA\ in the Gupta et al. (2012) sample, is 29. 
However two of the sources are our ``reference'' extragalactic targets Mkn~421 and PKS~2155-304, for which we here used all the 
data available to date, and repeated the reduction and analysis, and 
we do not include the blazar 3C~273 because its line of sight passes through a Supernova Remnant in our Galaxy.}
. 
The G-sample contains also the HRC-S-LETG spectrum of the blazar H~2356-309 ($z=0.165$), already published by Fang et al. (2010) and 
Zappacosta et al. (2010, 2012) and for which Buote et al. (2009) and Fang et al. (2010) proposed the detection of intervening OVII K$\alpha$ WHIM 
absorption in the redshift range $z\simeq 0.031-0.032$, where the Sculptor Wall intercepts the line of sight to this blazar (but see Nicastro et al., 2015, 
for a different interpretation: N15). 
For completeness we also include in our independent analysis the RGS spectrum of the blazar Mkn~501 ($z=0.0337$), not present in the original 
G-sample, but for which a similar claim of intervening OVII K$\alpha$ WHIM absorption, at a similar redshift of $z=0.0329 \pm 0.0008$ 
(consistent with the blazar's redshift of $z=0.0337$), has been recently proposed by the same team (Ren, Fang \& Buote, 2014, but see N15 for a 
different interpretation). 

Table 2 lists our 29 extragalactic targets (ordered by increasing 3$\sigma$ EW$_{thresh}$), together with: (a) their redshift (second column), 
(b) the line-of-sight column density of neutral H as inferred from 21-cm measurements (Kalberla et al., 2005; third column), and (c) 
as measured by us through spectral fitting of the HETG, LETG and RGS spectra (fourth column), (d) the instrumental configuration used in 
our spectral analysis, (e) the SNRE and (f) 3$\sigma$ EW$_{thresh}$ at 22.2 \AA\ of their spectra. 
\begin{table*} 
\begin{minipage}{126mm}
\caption{\bf Extragalactic Sources}
\begin{tabular}{|ccccccc|}
\hline
Source Name & Redshift & $^a$N$_H$ (21 cm) & N$_H^X$ & Spectrometer & SNRE$^b$ & 3$\sigma$ EW$_{Thresh}$ \\
& & in 10$^{20}$ cm$^{-2}$ & in 10$^{20}$ cm$^{-2}$ & & at 22.2 \AA\ & at 22.2 \AA, in m\AA\ \\ 
\hline
\multicolumn{7}{|c|} {Reference Targets} \\
Mkn~421 & 0.03002 & 1.9 & $1.9^{+0.1}$ & ACIS-S-LETG & 80.0 & 1.9 \\
Mkn~421 & 0.03002 & 1.9 & $1.9^{+0.2}$ & HRC-S-LETG & 74.0 & 2.0 \\
PKS~2155-304 & 0.11600 & 1.5 & $1.5^{+0.1}$ & ACIS-S-LETG & 60.0 & 2.5 \\
PKS~2155-304 & 0.11600 & 1.5 & $1.5^{+0.1}$ & HRC-S-LETG & 49.0 & 3.1 \\
\multicolumn{7}{|c|} {Gupta et al. (2012) Parent Sample} \\
Mkn~279 & 0.030405 & 1.5 & $1.5^{+0.7}$ & HRC-S-LETG & 18.7 & 8.0 \\
Mkn~509 & 0.03440 & 4.3 & $4.4_{-0.1p}^{+0.3}$ & ACIS-S-HETG & 6.8 & 11.0 \\
H~2356-309 & 0.165388 & 1.4 & $1.51 \pm 0.07$ & HRC-S-LETG & 12 & 12.5 \\
NGC~4593 & 0.00900 & 1.9 & $2.3_{-0.4p}^{+0.3}$ & ACIS-S-HETG & 4.4 & 17.0 \\
Mkn~290 & 0.03040 & 1.8 & $1.8^{+0.09}$ & ACIS-S-HETG & 4.0 & 18.8 \\
NGC~3783 & 0.00973 & 9.9 & $9.9^{+1.5}$ & ACIS-S-HETG & 4.0 & 18.8 \\
Ark~564 & 0.02467 & 5.3 & $7.3 \pm 1.0$ & ACIS-S-HETG & 4.0 & 18.8 \\
MR~2251-178 & 0.06398 & 2.4 & $2.4^{+0.1}$ & HRC-S-LETG & 7.7 & 19.5 \\
H~1426+428 & 0.12900 & 1.1 & $1.1^{+0.3}$ & ACIS-S-HETG & 3.2 & 23.4 \\
1ES~0120+340 & 0.272 & 5.2 & $14.1 \pm 0.4$ & HRC-S-LETG & 5.9 & 25.4 \\
Mkn~501 & 0.033663& 1.6 & $5.5 \pm 0.3$ & RGS & 7.5 & 28.0 \\
3C~454.3 & 0.85900 & 6.6 & $12.4 \pm 0.4$ & HRC-S-LETG & 5.3 & 28.3 \\
NGC~5548 & 0.01718 & 1.6 & $1.6^{+1.1}$ & ACIS-S-HETG & 2.4 & 31.3 \\
BL~0502+675 & 0.31400 & 9.1 & $6.2 \pm 0.3$ & HRC-S-LETG & 4.7 & 31.9 \\
Mkn~1044 & 0.16451 & 3.3 & $4.2 \pm 0.2$ & HRC-S-LETG & 4.7 & 31.9 \\
NGC~7469 & 0.01639 & 4.5 & $5.2_{-0.3p}^{+0.3}$ & ACIS-S-HETG & 1.9 & 39.5 \\
H~2106-099 & 0.02652 & 6.6 & $11.0_{-0.8}^{+0.9}$ & HRC-S-LETG & 3.5 & 42.9 \\
3C~382 & 0.05790 & 7.0 & $7.0^{+0.9}$ & ACIS-S-HETG & 1.7 & 44.1 \\
IRAS~13349+2438 & 0.10764 & 1.0 & $1.0^{+0.1}$ & ACIS-S-HETG & 1.7 & 44.1 \\
1ES~1028+511 & 0.36040 & 1.2 & $4.7 \pm 1.1$ & ACIS-S-HETG & 1.7 & 88.2 \\
NGC~3516 & 0.00884 & 3.5 & $16 \pm 3$ & ACIS-S-HETG & 1.6 & 46.9 \\
1ES~1927+654 & 0.01700 & 6.9 & $17 \pm 1$ & HRC-S-LETG & 3.2 & 46.9 \\
NGC~4051 & 0.00242 & 1.2 & $1.74 \pm 0.01$ & HRC-S-LETG & 2.8 & 53.6 \\
NGC~1275 & 0.01756 & 14 & $24 \pm 2$ & ACIS-S-HETG & 1.4 & 53.6 \\
PG 0844+349 & 0.06400 & 2.9 & $2.9^{+0.2}$ & ACIS-S-HETG & 1.2 & 62.5 \\ 
H~1821+643 & 0.29700 & 3.4 & $3.4^{+0.2}$ & ACIS-S-HETG & 1.2 & 62.5 \\
NGC~3227 & 0.00386 & 2.0 & $6_{-2}^{+4}$ & HRC-S-LETG & 1.9 & 78.9 \\
\hline
\end{tabular}
$^a$ Weighted average N$_H$ from the Karberla et al. (2005) survey. 

$^b$ We assume 70 m\AA, 50 m\AA\ and 25 m\AA\ for the widths of the RGS, LETG and HETG resolution elements, respectively. 
Note that this means different unresolved-line EW threshold sensitivities for a given SNRE in the three different spectrometers, 
with the HETG being $70/25=2.8$ times more sensitive than the RGS to unresolved lines. 
\end{minipage}
\end{table*}

\subsection{Spectral Fitting}
We used the fitting package {\em Sherpa}, in CIAO (Freeman, Doe \& Siemignowska, 1999) to perform spectral fitting of the RGS1 
(the only RGS available in the 21-24 \AA\ spectral band, where the features of interest to this work are found) and LETG/HETG spectra 
of our Galactic and extragalactic targets. 
Our two main objectives were: (a) obtaining robust estimates of the Galactic neutral hydrogen column density N$_H^X$ attenuating 
the source continuum along our lines of sight through bound-free photoelectric absorption by elements lighter than oxygen at 
$\lambda \gs 10$ \AA, and (b) estimating line-of-sight neutral and ionized oxygen column densities through equivalent-width (EW) 
measurements of the main (K$\alpha$ and K$\beta$) bound-bound OI-OVII resonant transitions imprinted by the intervening gas in the 
21-24 \AA\ portion of the spectra of our Galactic and extragalactic samples. 
\footnote{An important caveat here is that, while our estimates of $z=0$ absorption against the lines of sight to extragalactic sources can only refer 
to physical and kinematical parameters of Galactic diffuse gaseous components, estimates of the same parameters against Galactic X-ray 
binaries, could in principle be (at least partly) contaminated by possible contributions from intrinsic (to the binary) neutral, mildly ionized 
or hot absorbers. However, while variable (and so intrinsic) hot or moderately ionized absorbers have been detected in several X-ray binaries (e.g. 
Pinto et al., 2014; van Peet et al., 2009; Madej et al., 2010), evidence for neutral intrinsic or low-ionization absorbers is sparse and controversial 
(e.g. Schulz et al., 2010)}. 

To measure N$_H^X$, we first modeled the broad-band (10-30 \AA, from which we excluded the narrow regions 21.5-21.7 \AA\ and 
23.3-23.6 \AA, where the strong K$\alpha$ lines of OVII and OI and OII, respectively, are present) spectra of our sources with two different 
continuum models: for Galactic X-Ray Binaries (XRBs) we used a combination of black body (model {\em xsbbody} in {\em Sherpa}) and 
power-law (model {\em xspowerlaw} in {\em Sherpa}), while for our extragalactic AGN spectra we used a simple power-law. 
In all cases we attenuated the continua at long wavelengths through bound-free photoelectric absorption by neutral gas (the XSPEC-native 
model {\em xstbabs} in {\em Sherpa}: Wilms, Allen \& McCray, 2000). 
{\em xstbabs} includes photoelectric absorption by the bound-free K-edge transitions of neutral H, He and light metals but does not 
model the many bound-bound resonant K transitions from OI that populate the OI K trough between 22.6-23.5 \AA, most of which are 
not individually detected or resolved in RGS and LETG/HETG spectra (e.g. Gatuzz et al., 2013a). 
Moreover {\em xstbabs} does not account for absorption by mildly ionized oxygen (as well as other species), which also imprint both 
unresolved lines (OII-OVII) and edges (OII) in the explored 21-24 \AA\ spectral interval, as well as FeII L edges at 17.1-17.6 \AA, and 
therefore the column $N_H^X$ measured via {\em xstabs} is effectly an estimate of the HI column and generally underestimates the actual 
equivalent H column if the absorbing gas is not completely neutral. Despite these omissions, {\em xstbabs} describes reasonably well the 
observed trough in the 22.5-23.5 \AA\ region, both in our Galactic and extragalactic spectra. 

\begin{figure*} 
\begin{minipage}{126mm}
\begin{center}
\includegraphics[width=120mm]{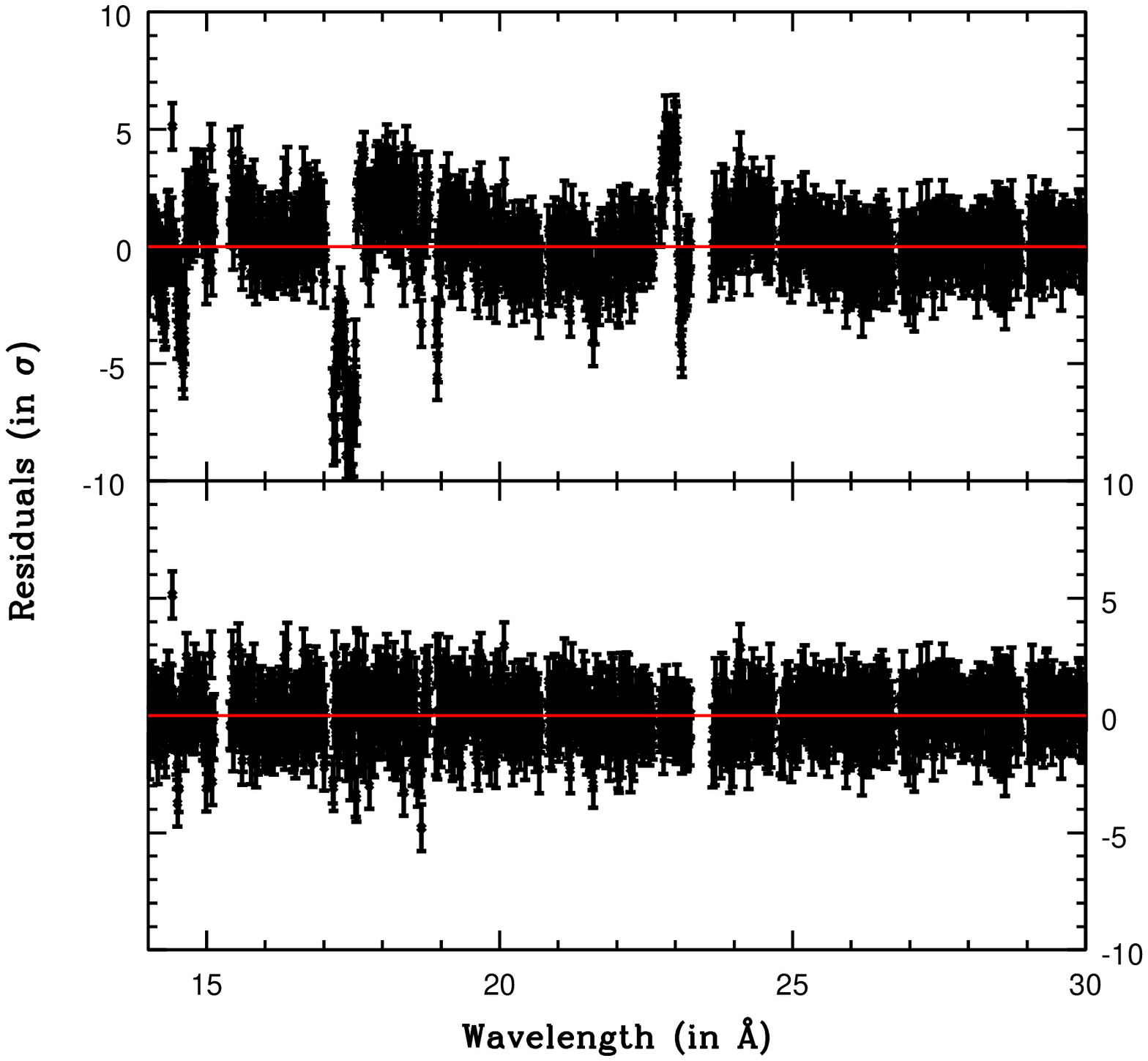}
\end{center}
\caption{Top panel: residuals to the best-fitting attenuated-continuum model of the 14-30 \AA\ spectrum of the XRB V*V~821 Ara, which 
suffered the strongest systematic deviations ($\chi^2_r(dof) = 3.2(1407)$). Bottom panel: residuals to the best-fitting attenuated-continuum 
plus Gaussians model of the 14-30 \AA\ spectrum of the XRB V*V~821 Ara, yielding $\chi^2_r(dof) = 1.08(1409)$.}
\label{res}
\end{minipage}
\end{figure*}

\noindent 
However, a visual inspection of the 10-30 \AA\ residuals often revealed systematic broad-band (and also narrow, in correspondence of the 
Fe II, L edges at 17.1-17.6 \AA\ and the OI K$\beta$, K$\gamma$ lines at 22.7-23.3 \AA, or simply due to moderate pile-up: see below) 
deviations (especially in the highest S/N spectra of XRBs), which were clearly reflected in the reduced $\chi^2$ that in a few cases had values 
as high as $\chi_r^2 \simeq 3$. 
Fig. \ref{res} (top panel) shows the residuals to the best-fitting attenuated-continuum model of the 14-30 \AA\ spectrum of the XRB 
V*V~821 Ara (the RGS1 is blind in the 10-14 \AA\ region), which suffered the strongest of such systematic deviations: $\chi^2_r(dof) = 
3.2(1407)$. 
The broand band systematic wiggling of these residuals is most likely a consequence of a poor intrinsic-continuum modeling: our high 
SNRE XRB (and to a less extent AGN) spectra are, each, the result of the sum of a number of source spectra taken in different epochs and 
with different intrinsic source continua (both in shape and normalization). The resulting continuum is therefore a complicated combination 
of power-laws with different spectral indeces and black body emission with different temperatures, which however we model with a 
combination of a single power-law plus a single-temperature black-body. 
Such mis-modeling of the source continua, could in principle yield to unreliable N$_H^X$ best-fitting values, unless the broad band 
deviations are opportunely cured by the inclusion in the model of additional components. To satisfactorily model these deviations, we 
then progressively added relatively broad (i.e. well-resolved: $\ge 3 \times$ the RGS, LETG or HETG resolution element) absorption or 
emission Gaussians, whenever and wherever needed, and repeated the fit till an acceptable 10-30 \AA\ value of the statistics was 
reached (i.e. $\chi^2_r = 0.8-1.2$). 
The bottom panel of Fig. \ref{res} shows the final result of these procedure, for the spectrum of V*V~821 Ara ($\chi^2_r(dof) = 1.08(1409)$). 
This procedure yields accurate descriptions of the 10-30 \AA\ spectra, but could also in principle still affect in a systematic way our 
measurements of N$_H^X$, which are vital to estimate the metallicity of the neutral to mildly ionized absorbing medium (see below). 
To check that our final N$_H^X$ measurements are indeed reliable, we compared the best-fitting N$_H^X$ value that we obtain for this 
parameter by fitting, through the above described procedure, the total co-added spectrum of one of the most variable XRB of our sample, 
Cygnus X-1, with that that we obtain by fitting simultaneously the six RGS1 spectra of this source collected during six different observations, 
each with its own continuum. 
We find: N$_H^X = (5.3 \pm 0.1)\times 10^{21}$ cm$^{-2}$, in the fit to the total coadded spectrum, and N$_H^X = (6.1 \pm 0.8)\times 10^{21}$ 
cm$-^{2}$,  in the simultaneous fits to the 6 spectra of Cygnus X1. These two values are fully consistent with each other within their 
$1\sigma$ statistical errors, confirming the reliability of our N$_H^X$ (Table 1 and 2, columns 3 and 4, respectively). 

Finally, for the spectra of the XRBs of our sample (that can reach in a few cases 6-30 A fluxes of few tenths of Crab), another problem that 
can in principle affect our N$_H^X$ measurements is that of pile-up.  
The effect of pile-up in RGS spectra is twofold: (a) it causes a migration of 1$^{st}$ order long-wavelength photons to 2$^{nd}$ order 
short-wavelength (half) photons, and (b) it introduces spurious narrow features in the 1$^{st}$ order spectrum, due to the dependence of 
pile-up on the exact location of the CCD that is hit by the piled-up photons. 
The second effect might result in slightly poorer modeling of the broad-band continuum and subsequent slightly worse $\chi^2$, 
and this effect, if present, is cured through the procedure described above. 
The first effect, however, if seriously present, could in principle artificially modify the curvature of the soft X-ray spectral shape produced 
by photo-electric neutral absorption and therefore falsify the estimate of N$_H^X$. 
Fortunately, for  virtually all the XRBs of our sample, we dispose of more than one spectrum. And in almost all the cases potentially affected by 
pile-up in the co-added spectrum, we have spectra with high count rate (causing moderate pile-up) and spectra with low count-rate 
(unaffected by pile-up). 
To estimate the degree of pile-up in our high-flux, high SNRE, XRB spectra and evaluate its effect on our N$_H^X$ measurements, 
we first ratioed the first- and second-order RGS1 spectra and then, for those sources that showed deviations from unity in this ratio, 
we modeled the high- and low-state spectra separately, with continuum models attenuated by {\em xstbabs}, and estimated the 
differences in measured N$_H^X$. 
Our conlcusion is that none of the total coadded spectra of our XRBs are affected by pile-up by more than 10\% and in all the cases 
studied the degree of pile-up is not sufficient to affect significantly our N$_H^X$ measurement. 
We conclude that pile-up is not an important issue for the XRB spectra of our sample, and so for our analysis.

We then proceded to model the possible presence of narrow $z=0$ bound-bound K absorption from neutral and ionized oxygen (or to set 
EW upper limits for these transitions), locally, in the 21-24 \AA\ band. 
In particular, we modeled our 21-24 \AA\ spectra with local continua attenuated by the 10-30 \AA\ best-fitting {\em xstbabs} models 
(to proper model the OI K edge), and added 7 unresolved (line FWHM frozen to 10 m\AA\ and intrinsically broadened to the instrument 
resolution width through the convolution product with the instrument Line Response Function) negative Gaussians to our best-fitting 
continuum models, at the rest frame wavelengths of the: (1) OI K$\alpha$ (e.g. Juett et al., 2004), (2) OII K$\alpha$ (e.g. Juett et al., 2004); 
(3) OII K$\beta$ (Gatuzz et al., 2013a,b); (4) OIII-OVI K$\alpha$ (e.g. Garcia et al., 2005); and (5) OVII K$\alpha$ (e.g. Verner, Verner \& 
Ferland, 1996), and repeated the fit by allowing all line normalizations to vary freely and their centroids to vary around the respective rest 
frame positions by $\pm 1$ spectrometer resolution element ($\pm 70$ m\AA\ for the RGS, $\pm 50$ m\AA\ for the LETG and 
$\pm 30$ m\AA\ for the HETG). 

\begin{figure*} 
\begin{minipage}{126mm}
\begin{center}
\includegraphics[width=120mm]{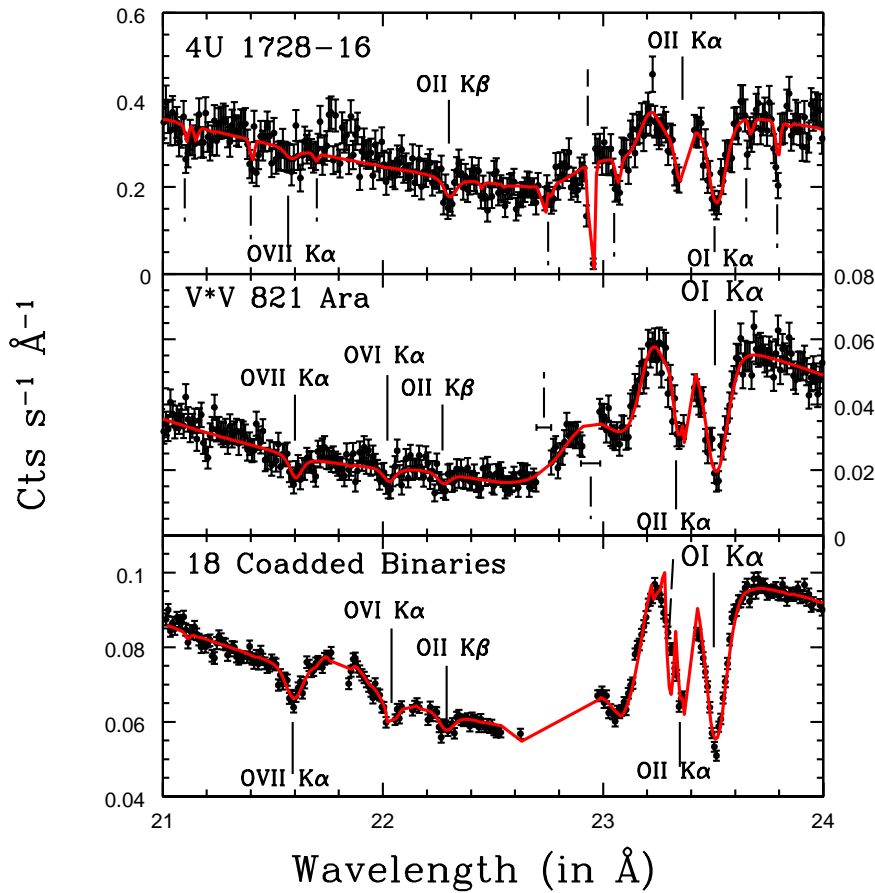}
\end{center}
\caption{ 21-24 \AA\ portions of the high SNRE RGS spectra of the X-ray binary 4U~1728-17 (top panel), 
the X-ray binary V*V~821 Ara (middle panel) and the coadded remaining 18 X-ray binaries of our 
Galactic sample. Long/Short-dashed vertical lines mark the positions of RGS instrumental features.}
\label{galspec}
\end{minipage}
\end{figure*}

\noindent
Figure \ref{galspec} shows the three highest-SNRE examples of the presence of some of these lines in the spectra of the X-ray binaries 
4U~1728-16 (Fig. \ref{galspec}, top panel), V*V~821 Ara (Fig. \ref{galspec}, middle panel), and the coadded spectrum of the remaining 
18 X-ray binaries of our Galactic sample (Fig. \ref{galspec}, bottom panel). 

\begin{figure*} 
\begin{minipage}{126mm}
\begin{center}
\includegraphics[width=120mm]{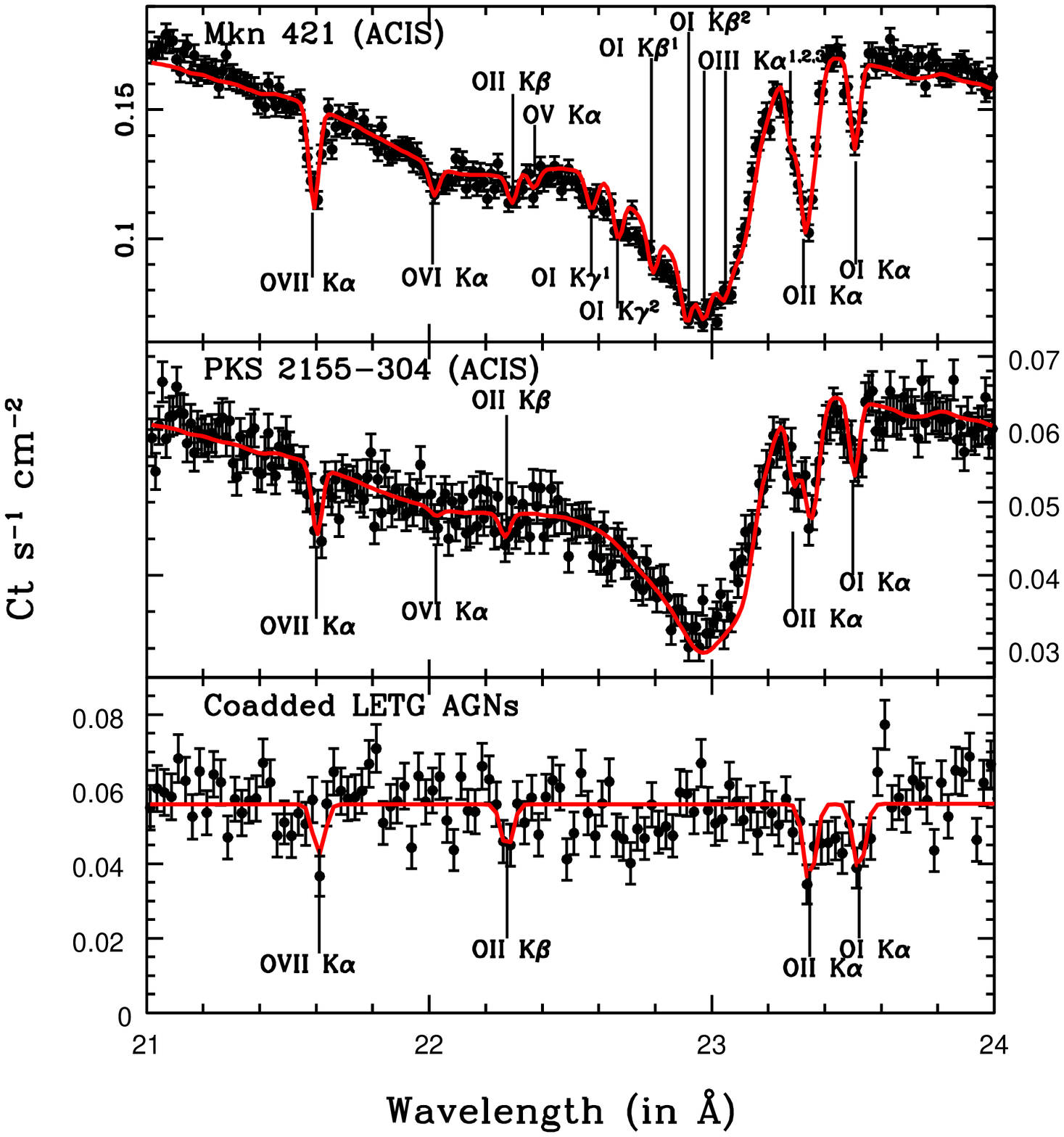}
\end{center}
\caption{ 21-24 \AA\ portions of the high SNRE LETG spectra of the blazars Mkn~421 (top panel), 
PKS~2155-309 (middle panel) and the coadded remaining 8 AGNs of our extragalactic sample, observed with the {\em 
Chandra}-LETG and not contaminated by intrinsic absorption/emission at the relevant wavelengths (see Table 4 and \S 2.5 for details).}
\label{exgalspec}
\end{minipage}
\end{figure*}

\noindent
Figure \ref{exgalspec} shows three analogous examples for our extragalactic sources. 

For the lines that were succesfully modeled by our Gaussians, we proceeded to determine their EWs (with the routine {\em eqwidth} in
{\em Sherpa}) and then evaluated the $1\sigma$ statistical uncertainties on their positions and EWs (through the {\em Sherpa} 
routine {\em projection}). 
Whenever, instead, one of these lines was not found by the fitting procedure (i.e. a null best-fit normalization), we froze the line position to 
its rest-frame value and used the {\em Sherpa} {\em projection} routine to evaluate its 3$\sigma$ EW upper limit. 

\noindent
Here we only report on K-shell transitions from neutral and mildly ionized O, in particular OI K$\alpha$, OII K$\alpha$ (where possible: see 
\S 2.4) and OII K$\beta$. 
We do not model the two OI K$\beta$ transitions reported by Gatuzz et al. (2013a,b) because these lines falls within the deep OI K edge trough, 
just leftward the edge wavelength of $\simeq 23.31$ \AA, where the S/N of our data (with the only exception of the ACIS-LETG spectrum 
of Mkn~421: Fig. \ref{exgalspec}, top panel) is not sufficient to perform this study. 

\noindent
A discussion of the higher ionization, $z=0$ component is deferred to forthcoming papers. 

\subsection{Results for the Galactic Sample}
Table 3 summarizes the results of our search for neutral (OI) and low-ionization (OII) bound-bound absorption, for the 20 sources 
of our Galactic sample. Targets are ordered by decreasing unresolved absorption line sensitivity of their spectra at 22.2 \AA.  
For all the reported lines, the uncertainty on their positions corresponds to half RGS resolution element of 30 m\AA. 

Strong (EW$=12-110$ m\AA) OI K$\alpha$ absorption is detected along all Galactic lines of sight, at a mean wavelength 
$<\lambda> = 23.52 \pm 0.04$  (the associated uncertainty is the maximum semi-dispersion; all later mean values will use the 
same measure of uncertainty, unless explicitly stated), with statistical significances ranging between 2.4-36$\sigma$ (only in one case 
is the statistical significance $\le 3\sigma$). 

The RGS spectra of 12 out of the 20 Galactic lines of sight are blocked in the 23.3-23.4 \AA\ interval by the presence of an instrumental 
feature due to the correspondence between cool pixels in the read-out detector and these dispersed wavelengths. 
This spectral interval is also where the rest frame position of the OII K$\alpha$ absorption line falls ($\lambda($OII K$\alpha) = 23.35$ \AA\, 
e.g. Juett et al., 2004). 
This unfortunate circumstance makes it impossible to properly model the OII K$\alpha$ absorption line along these 12 lines of sight (Table 3). 
However, in each of the remaining 8 spectra, a strong (EW$=20-87$ m\AA) OII K$\alpha$ absorption line is clearly detected at 
$<\lambda> = 23.35 \pm 0.03$, with statistical significances ranging between 4-33 $\sigma$. 

Of these 8 Galactic lines of sight for which we detect OII K$\alpha$, 6 also show indications for the associated OII K$\beta$ absorption 
line, near 22.3 \AA, previously identified by Gatuzz et al. (2013a,b) . 
The 2 sources for which we detect OII K$\alpha$ and that do not show evidence for the OII K$\beta$ line, are Cygnus~X-1 and Cygnus~X-2, 
for which we measure 3$\sigma$ upper limit of EW$<15$ and 9 m\AA, respectively. 
The OII K$\beta$ line is also hinted at in the spectra of three additional Galactic binaries for which the OII K$\alpha$ line detection is hampered 
by the presence of the RGS instrumental feature at 23.3-23.4 \AA. 
This makes the total number of possible OII K$\beta$ absorption lines along the Galactic lines of sight, equal to 9 out of the 20 lines of 
sight inspected, with statistical significances ranging between 1.1-3.7$\sigma$. 
The average position of the OII K$\beta$ line in the RGS spectra of these 9 Galactic sources is $<\lambda> = 22.29 \pm 0.04$ \AA. 

\begin{table*} 
\begin{minipage}{126mm}
\caption{\bf Bound-Bound OI and OII K Absorption for the Galactic Sample$^a$}
\begin{tabular}{|ccccccc|}
\hline
Source Name & $\lambda$(OI K$\alpha$) & EW(OI K$\alpha$) & $\lambda$(OII K$\alpha$) & EW (OII K$\alpha$) & $\lambda$(OII K$\beta$) & EW(OII K$\beta$) \\
& in \AA\ & in m\AA\ & in \AA\ & in m\AA\ & in \AA\ & in m\AA\ \\ 
\hline
Her~X-1 & $23.52 \pm 0.03$ & $16 \pm 2$ & N/A$^{b}$ & N/A$^{b}$ &  22.29$^{c}$ & $<7$ \\ 
EXO~0748-676 & $23.52 \pm 0.03$ & $52 \pm 3$ & N/A$^{b}$ & N/A$^{b}$ &  22.29$^{c}$ & $<9$ \\
PSRB~0833-45 & $23.56 \pm 0.03$ & $12 \pm 5$ & N/A$^{b}$ & N/A$^{b}$ &  $22.29 \pm 0.03$ & $8 \pm 5$ \\
SAX~J1808.4-3658 & $23.52 \pm 0.03$ & $59 \pm 3$ & N/A$^{b}$ & N/A$^{b}$ & $22.30 \pm 0.03$ & $9 \pm 3$ \\
SWIFT~J1753.5-0127 & $23.51 \pm 0.03$ & $72 \pm 2$ & N/A$^{b}$ & N/A$^{b}$ & 22.29$^{c}$ & $<10$ \\
Cygnus~X-2 & $23.48 \pm 0.03$ & $77 \pm 3$ & $23.33 \pm 0.03$ & $66 \pm 2$ & 22.29$^c$ & $<9$ \\
MAXI~J0556-332 & $23.50 \pm 0.03$ & $19 \pm 5$ & $23.33 \pm 0.03$ & $20 \pm 5$ &  $22.32 \pm 0.03$ & $4.3 \pm 3.8$ \\
Cygnus~X-1 & $23.51 \pm 0.03$ & $107 \pm 3$ & $23.36 \pm 0.03$ & $40 \pm 2$&  22.29$^{c}$ & $<15$ \\
SWIFT~J1910.2-0546 & $23.52 \pm 0.03$ & $66 \pm 3$ & $23.38 \pm 0.03$ & $41 \pm 3$ & $22.27\pm 0.03$ & $7 \pm 4$ \\
4U~1636-53 & $23.51 \pm 0.03$ & $94 \pm 5$ & $23.33 \pm 0.03$ & $87 \pm 6$ & $22.29 \pm 0.03$ & $9 \pm 6$ \\
4U~1728-16 & $23.51 \pm 0.03$ & $76 \pm 5$ & $23.34 \pm 0.03$ & $40\pm 4$ & $22.30\pm 0.03$ & $22 \pm 6$ \\
V*V821Ara & $23.51 \pm 0.03$ & $110 \pm 5$ & $23.33 \pm 0.03$ & $56 \pm 2$ & $22.27 \pm 0.03$ & $19 \pm 6$ \\
GS~1826-238 & $23.51 \pm 0.03$ & $63 \pm 4$ & N/A$^{b}$ & N/A$^{b}$ &  22.29$^{c}$ & $<25$ \\
HETE~J1900.1-2455 & $23.52 \pm 0.03$ & $51 \pm 6$ & N/A$^{b}$ & N/A$^{b}$ &  22.29$^{c}$ & $<23$ \\
4U~2129+12 & $23.55 \pm 0.03$ & $25 \pm 8$ & N/A$^{b}$ & N/A$^{b}$ &  22.29$^{c}$ & $<28$ \\
4U~1543-624 & $23.51 \pm 0.03$ & $100 \pm 6$ & $23.36 \pm 0.03$ & $78 \pm 5$ & $22.31 \pm 0.03$ & $20 \pm 9$ \\
Aql~X-1 & $23.51 \pm 0.03$ & $75 \pm 4$ & N/A$^{b}$ & N/A$^{b}$ & $22.25 \pm 0.03$ & $29 \pm 11$ \\
4U~1735-444 & $23.51 \pm 0.03$ & $79 \pm 9$ & N/A$^b$ & N/A$^b$ & 22.29$^c$ & $<35$ \\
X~Persei & $23.52 \pm 0.03$ & $68 \pm 12$ & N/A$^{b}$ & N/A$^{b}$ &  22.29$^{c}$ & $<43$ \\
XTE~J1650-500 & $23.52 \pm 0.03$ & $101 \pm 10$ & N/A$^{b}$ & N/A$^{b}$ & 22.29$^c$ & $<61$ \\
\hline
18CB & $23.50 \pm 0.03$ & $74.4 \pm 0.8$ & $23.35 \pm 0.03$ & $41.8 \pm 0.6$ & $22.28 \pm 0.03$ & $9.2 \pm 1.3$ \\
\hline
\end{tabular}
Uncertainties on line positions are given in half RGS resolution element. 

$^a$ We give EW values for all measurements, and 3$\sigma$ upper limits for the rest

$^b$ Overlapping RGS Instrumental Feature. 

$^c$ Line position frozen to evaluate the EW $3\sigma$ upper limit. 
\end{minipage}
\end{table*}

\subsection{Results for the Extragalactic Sample}
Table 4 summarizes the results of our search for neutral (OI) and low-ionization (OII) bound-bound absorption, for the 29 sources (31 spectra) of our 
extragalactic sample. Targets are ordered by decreasing unresolved absorption line sensitivity of their spectra at 22.2 \AA. 
Wavelength accuracies in Table 4 correspond to half the width of the spectrometer resolution element (30 m\AA\ for the LETG and the 
RGS and 10 m\AA\ for the HETG). 
For our two reference extragalactic targets, Mkn~421 and PKS~2155-304, for which 2 LETG spectra are available, the measured EWs of 
the OI and OII lines are always consistent with each other in the two spectra, within their 90\% errors. For these two lines of sight, in Table 4 
we therefore report the averages of the EWs measured in each of the two spectra, together with their associated propagated (in quadrature) 
1$\sigma$ uncertainties. 

Also in our extragalactic sample, we found ubiquitous evidence for the presence of relatively strong (EW$=7-50$ m\AA) OI K$\alpha$ 
absorption, with statistical significances in the range 1.1-21$\sigma$. 
The mean wavelength of the OI K$\alpha$ line in our extragalactic sample, is $<\lambda> = 23.51 \pm 0.05$. 
The OI K$\alpha$ line is not hinted at in only 3 out of the 31 spectra of our sample, that of MR~2251-178 (heavily contaminated by at 
least two intrinsic warm absorber components) and those of PG~0844+349, and NGC~3227 (among those with the lowest SNRE in our 
extragalactic sample: Table 2). 
For these three lines of sight we measure 3$\sigma$ upper limits of EW(OI K$\alpha$)$< 25$, 93 and 133 m\AA, respectively.  

Moderately strong (EW$=6-43$ m\AA) OII K$\alpha$ absorption is also seen in the vast majority (27 out of 31) of the spectra of our 
extragalactic sample, at a mean wavelength $<\lambda> = 23.35 \pm 0.05$ \AA\ and with statistical significances in the range 
1.3-15.6$\sigma$. 
Three out of the 4 cases in which the OII K$\alpha$ line is not hinted at in the data (the lines of sight toward 1ES~1927+654, NGC~1275 
and NGC~3227), have very low-SNRE spectra (Table 2) and we measure loose 3$\sigma$ upper limits of $<134$, 117 and 126 m\AA. 
In the fourth case, Mkn~501, the detection of the OII K$\alpha$ line is hampered by the presence of an instrumental feature in its RGS 
spectrum. 

Finally, 7 out of the 25 extragalactic lines of sight (27 spectra) that show the presence of OII K$\alpha$, also hint at 
an absorption line at an average wavelength of $<\lambda> =  22.28 \pm 0.02$, with statistical significances in the range 1-5.5$\sigma$. 
An absorption line at a consistent wavelength ($\lambda = 22.30 \pm 0.03$ is also clearly (3.5$\sigma$) seen in the RGS spectrum of 
the blazar Mkn~501, where the K$\alpha$ transition is blocked by an instrumental feature. 
Following Gatuzz et al. (2013a,b), and following the same procedure as for our Galactic sample, we identify this line as the K$\beta$ transition 
of OII, associated to the stronger K$\alpha$ transition that we also consistently detect in all the correspnding extragalactic spectra where 
the K$\alpha$ detection is possible. 
\noindent 
Twenty one spectra of our extragalactic sample lines of sight do not show the presence of the OII K$\beta$ transition. 
In 9 of them the detection of this line is hampered by the superimposition of intrinsic AGN absorption or emission (Table 4). 
For the remaining 12, the loose 3$\sigma$ EW upper limits that we measure at the OII K$\beta$ rest-frame wavelength are 
in the range 36-300 m\AA, fully consistent with the values reported for the 10 spectra in which this line is modeled (Table 4). 

\begin{table*} 
\begin{minipage}{126mm}
\caption{\bf Bound-Bound OI and OII K Absorption for the Extragalactic Sample$^a$}
\begin{tabular}{|ccccccc|}
\hline
Source Name & $\lambda$(OI K$\alpha$) & EW(OI K$\alpha$) & $\lambda$(OII K$\alpha$) & EW (OII K$\alpha$) & $\lambda$(OII K$\beta$) & EW(OII K$\beta$) \\
& in \AA\ & in m\AA\ & in \AA\ & in m\AA\ & in \AA\ & in m\AA\ \\ 
\hline
\multicolumn{7}{|c|}{Reference Targets} \\
$<$Mkn~421$>$ & $23.52 \pm 0.03$ & $8.5 \pm 0.5$ & $23.32 \pm 0.03$ & $9.9\pm 0.6$ &  $22.29 \pm 0.03$ & $3.3\pm 0.6$ \\
$<$PKS~2155-304$>$ & $23.51 \pm 0.03$ & $9.8 \pm 1.0$ & $23.32 \pm 0.03$ & $8.6\pm 1.1$ &  $22.29 \pm 0.03$ & $2.9\pm 1.1$ \\
\multicolumn{7}{|c|}{Gupta et al. (2012) Parent Sample} \\
Mrk~279 &  $23.50 \pm 0.03$ & $23 \pm 6$ & $23.34 \pm 0.03$ & $22 \pm 7$ & $^{c}$N/A  & $^c$N/A \\
Mkn~509 & $23.52 \pm 0.01$ & $16 \pm 6$ & $^b$$23.35 \pm 0.01$ & $^b$$33 \pm 8$ & $^{c}$N/A & $^c$N/A \\
H~2356-309 & $23.50 \pm 0.03$ & $19 \pm 9$ & $23.35 \pm 0.03$ & $27 \pm 11$ &  $22.28 \pm 0.03$ & $17 \pm  9$ \\
NGC~4593 & $23.51 \pm 0.01$ & $13 \pm 7$ & $^b$$23.33 \pm 0.01$ & $^b$$24 \pm 12$ & $^d$N/A & $^d$N/A \\
Mkn~290 & $23.51 \pm 0.01$ & $19 \pm 9$ & $^b$$23.34 \pm 0.01$ & $^b$$27 \pm 12$ &  $^{c}$N/A & $^c$N/A \\
NGC~3783 & $23.51 \pm 0.01$ & $26 \pm 11$ & $^b$$23.34 \pm 0.01$ & $^b$$27 \pm 12$ & $^d$N/A & $^d$N/A \\
Ark~564 & $23.50 \pm 0.01$ & $25 \pm 20$ & $^b$$23.35 \pm 0.01$ & $^b$$26 \pm 15$ & $^{c}$N/A & $^c$N/A \\
MR~2251-178 & $^{f}$23.51 & $<33$ & $23.39 \pm 0.03$ & $24\pm 8$ &  $^{c}$N/A & $^c$N/A \\
H~1426+428 & $23.51 \pm 0.01$ & $26 \pm 13$ & $^b$$23.34 \pm 0.01$ & $^b$$20 \pm 12$ & $^f$22.28 & $<59$ \\
1ES~0120+340 & $23.54 \pm 0.03$ & $44 \pm 20.0$ & $23.36 \pm 0.03$ & $28\pm 14$ &  $22.30 \pm 0.03$ & $23\pm 11$ \\
Mkn~501 & $23.51 \pm 0.03$ & $26 \pm 6$ & $^e$N/A & $^e$N/A & $22.30 \pm 0.03$ & $21 \pm 6$ \\
3C~454.3 & $23.48 \pm 0.03$ & $20 \pm 10$ & $23.37 \pm 0.03$ & $30\pm 10$ &  $22.27 \pm 0.03$ & $23\pm 9$ \\
NGC~5548 & $23.51 \pm 0.01$ & $17 \pm 6$ & $^b$$23.34 \pm 0.01$ & $^b$$32 \pm 4$ & $^{f,g}$22.28 & $^g$$<36$ \\
BL~0502+675 & $23.45 \pm 0.03$ & $34 \pm 14$ & $23.34 \pm 0.03$ & $37 \pm 15$ & $^f$22.28 & $<49$ \\
Mkn~1044 & $23.54 \pm 0.03$ & $24 \pm 11$ & $23.37 \pm 0.03$ & $31 \pm 11$ & $^f$22.28 & $<39$ \\
NGC~7469 & $23.49 \pm 0.01$ & $27 \pm 8$ & $^b$$23.34 \pm 0.01$ & $^b$$32 \pm 7$ & $^{f,g}$22.28 & $^g$$<71$  \\
H~2106-099 & $23.51 \pm 0.03$ & $21 \pm 20$ & $23.36 \pm 0.03$ & $34\pm 18$ &  $22.26 \pm 0.03$ & $34\pm 18$ \\
3C~382 & $23.53 \pm 0.01$ & $12 \pm 9$ & $^b$$23.34 \pm 0.01$ & $^b$$33 \pm 13$ & $^f$22.28 & $<53$ \\
IRAS~13349+2438 & $23.50 \pm 0.01$ & $13 \pm 12$ & $^b$$23.36 \pm 0.01$ & $^b$$21 \pm 11$ & $^f$22.28 & $<56$ \\
1ES~1028+511 & $23.51 \pm 0.01$ & $36 \pm 20$ & $^b$$23.34 \pm 0.01$ & $^b$$33 \pm 21$ &  $^f$22.28 & $<63$\\
NGC~3516 & $23.51 \pm 0.01$ & $25 \pm 4$ & $^b$$23.37 \pm 0.01$ & $^b$$27 \pm 3$ & $^{d}$N/A & $^d$N/A \\
1ES~1927+654 & $23.54 \pm 0.03$ & $34 \pm 18$ & $^f$23.35 & $<134$ & $^f$$22.28$ & $<80$ \\
NGC~4051 & $23.49 \pm 0.03$ & $19 \pm 7$ & $23.35 \pm 0.03$ & $26 \pm 5$ & $^{d}$N/A & $^d$N/A \\
NGC~1275 & $23.50 \pm 0.01$ & $50 \pm 33$ & $^{b,f}$23.35 & $^{b}$$<117$ & $^f$22.28 & $<233$ \\
PG 0844+349 & $^{f}$23.51 & $<93$ & $^b$$23.34 \pm 0.01$ & $^b$$43 \pm 33$ & $^f$22.28 & $<300$ \\
H~1821+643 & $23.50 \pm 0.01$ & $32 \pm 20$ & $^b$$23.34 \pm 0.01$ & $^b$$27 \pm 15$ & $22.29 \pm 0.01$ & $26 \pm 25$ \\
NGC~3227 & $^f$23.51 & $<133$ & $^f$23.35 & $^b$$<126$ & $^f$22.28 & $<120$ \\
\hline
8CLETG & $23.52 \pm 0.03$ & $17 \pm 4$ & $23.35 \pm 0.03$ & $20 \pm 4$ & $22.28 \pm 0.03$ & $11 \pm 4$ \\
9CHETG & $23.51 \pm 0.01$ & $20 \pm 11$ & $23.34 \pm 0.01$ & $27 \pm 12$ & $^f$22.29 & $<28$ \\
\hline
\end{tabular}
$^a$ We give EW values for all measurements, and 3$\sigma$ upper limits for the rest. 
$^b$ Affected by the OI in molecular form, due to the contaminant deposited on the ACIS window (Marshall et al., 2004).  
$^c$ Overlapping intrinsic OVII or OVIII K$\alpha$ Resonant absorption. 
$^d$ Overlapping intrinsic OVII K$\alpha$ Forbidden emission. 
$^e$ Overlapping RGS Instrumental Feature. 
$^f$ Line position frozen to evaluate the EW $3\sigma$ upper limit. 
$^g$ Nearby Instrinisc OVII K$\alpha$ Forbidden and Intercombination Emission. 
\end{minipage}
\end{table*}

\section{Discussion}
\begin{figure} 
\begin{center}
\includegraphics[width=84mm]{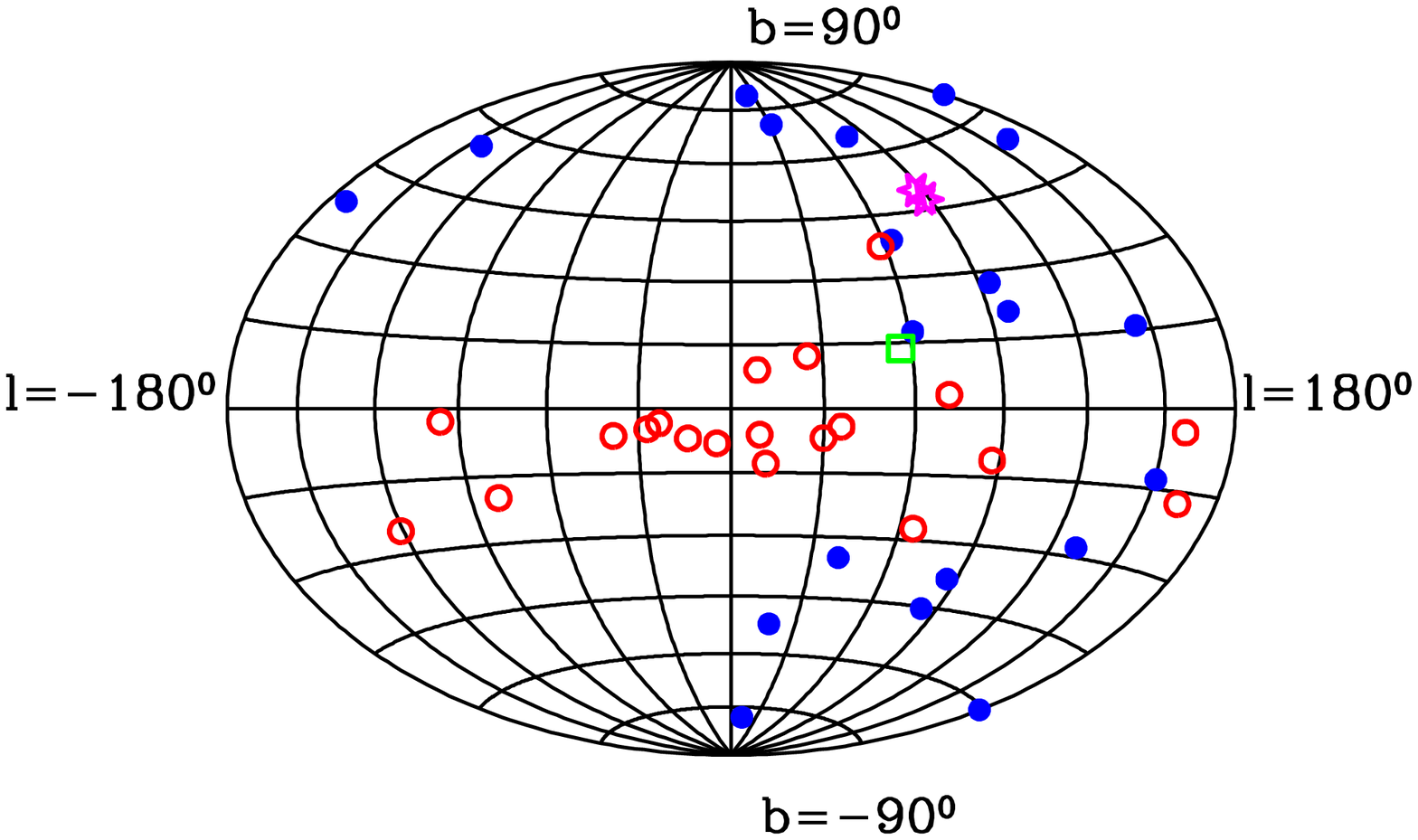}
\end{center}
\caption{ Aitoff projection map of the Galactic Coordinates of the targets of our Galactic (red open circles) and extragalactic (only the 20 targets that do not 
show contamination by intrinsic warm absorber/emitter components at the relevant wavelengths of our OI and OII transitions: blue solid circles) samples. 
The two starred magenta points show the S/N-weighted average directions of our extragalactic co-added spectra, while the green open square shows the 
S/N-weighted average direction of our Galactic co-added spectrum (see \S 3.2 and 3.3).}
\label{aitoff}
\end{figure}

Figure \ref{aitoff} shows the aitoff projection map (in Galactic coordinates, with $(l,b)=(0,0)$ in the center) of the targets of our Galactic (red open circles) and 
extragalactic (only the 20 targets that do not show contamination by intrinsic warm absorber/emitter components at the relevant wavelengths of our OI and OII 
transitions: blue solid circles) samples. 
The two starred magenta points show the "S/N-wieghted average directions" of our extragalactic co-added spectra, while the green open square shows the 
S/N-weighted average direction of our Galactic co-added spectrum (see \S 3.2 and 3.3). 
To preserve symmetry with respect to the Galactic plane (i.e. in Galactic latitudes $b$) and to the center versus anti-center (i.e. in Galactic 
longitudes $\ell$), these averages are computed on the absolute values of $b$ over $|b|\le 90^0$ and on the quantity $\Lambda$, over $\Lambda<180^0$, 
where $\Lambda$ is defined as $\Lambda = l$ for $0 \le l \le 180^0$ and $\Lambda = 360^0 - l$ for $180 < l < 360$. 
The majority (15 out of 20) of the lines of sight to Galactic sources are confined within latitudes $|b| \ls 15^{0}$ and therefore sample mainly the Galactic disk. 
Moreover, more than half of the sample (12 out of 20) lies in the direction of the Galactic center ($ 330^{0} \ls l \ls 30^{0}$) and only 
2 targets sample the anti-center direction ($150^{0} < l < 210^{0}$). 

On the contrary, all targets of our extragalactic sample, have Galactic latitudes $|b| \ge 15^{0}$, and therefore sample smaller portions of the 
Galactic disk, and are potentially able to probe material in the Galactic halo, or extended corona or even at larger distances, in the circum-galactic 
medium (CGM). 

\subsection{Strength of OI and OII Absorption in the two Samples: two distinct Absorbers?}
\begin{figure} 
\begin{center}
\includegraphics[width=84mm]{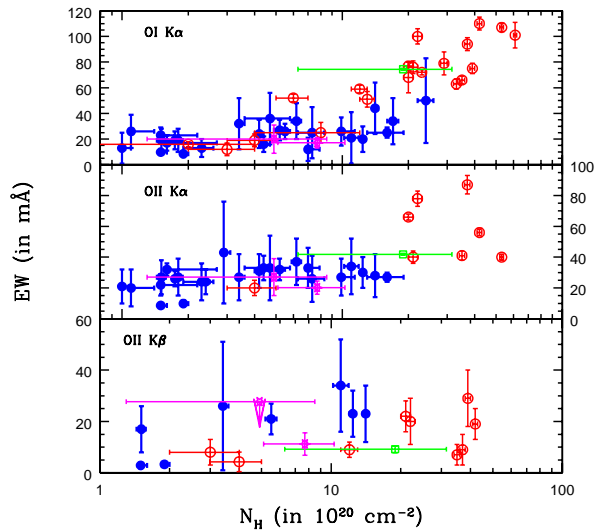}
\end{center}
\caption{EWs of OI K$\alpha$ (top panel), OII K$\alpha$ (middle panel) and OII K$\beta$ (bottom panel) as a function of the X-ray measured column of 
neutral gas N$_H^X$. Blue solid circles refer to Extragalactic targets, while red empty circles refer to Galactic XRBs. The green open square and the two magenta 
starred points refer to the measurements in the coadded spectra of Galactic and extragalactic sources, respectively: see \S 3.2, 3.3, last row of Table 3 and last 
two rows of Table 4).}
\label{ewoikaoiikaoiikbvsnh}
\end{figure}

OI and OII absorption is detected towards lines of sight to both Galactic and extragalactic sources. 
However, the strength and general characteristics of these absorbers are different in the two samples. 
Figure \ref{ewoikaoiikaoiikbvsnh} shows the measured EWs of the OI K$\alpha$ (top panel), OII K$\alpha$ (middle panel) and OII K$\beta$ (bottom panel), 
respectively, as a function of the estimated column of neutral gas N$_H^X$ as inferred from the gentle curvature of the X-ray continuum at energies $E \ls 2$ 
keV, due to bound-free opacity from neutral gas. In each panel, the red open circles refer to measurements along lines of sight to Galactic targets, while blue 
solid circles are relative to the extragalactic sources (the green open square and the two magenta starred points refer to the measurements in the coadded spectra 
of Galactic and extragalactic sources, respectively: see \S 3.2, 3.3, last row of Table 3 and last two rows of Table 4). 
The EWs of the OI K$\alpha$ along the Galactic lines of sight correlate strongly with N$_H^X$, indicating not only that the same gas responsible for the attenuation 
of the X-ray continuum at $E\ls 2$ keV is also imprinting the bound-bound OI transition in the X-ray spectra of the Galactic X-ray binaries, but also that 
such OI K$\alpha$ absorption is only moderately saturated. 
On the contrary, the OI K$\alpha$ EWs measured along the extragalactic lines of sight are not only, on average, smaller than those measured along our Galactic 
sightlines, but also virtually independent on N$_H^X$, over more than an order of magnitude in N$_H^X$, suggesting that two independent and physically distinct 
absorbers are responsible for the bulk of the OI K$\alpha$ and N$_H^X$ measurements along the extragalactic lines of sight: 
one asociated with the cold material in the portion of Galactic disk crossed by the lines of sight and producing the gentle curvature of the soft X-ray continuum 
and the other associated with diferent material at larger distances, producing most of the OI K$\alpha$ absorption.  

The OII K$\alpha$ and K$\beta$ absorbers show similar trends, but with lower statistical significances due to the smaller number and lower statistical significance of 
the OII K$\alpha$ and K$\beta$ absorbers, compared to OI K$\alpha$. 
To increase the statistical significance of our detections and so infer the fundamental properties of the likely different materials that imprint OI and OII 
absorption along the Galactic and extragalactic lines of sight, we decided to co-add some of our Galactic and extra-galactic spectra, according to the 
following criteria. 

\subsection{The Coadded RGS Spectrum of the Galactic X-Ray Binaries}
For the Galactic lines of sight, the EWs of the OII K$\alpha$ transitions (where detectable) are generally significantly larger than those of the 
corresponding K$\beta$ transitions (Fig. \ref{ewoikaoiikaoiikbvsnh}, middle and bottom panels), possibly indicating moderate saturation (see \S 3.4), and 
consequently so are the associated statistical significances of the detections. 
Indeed, only 3 of the 9 OII K$\beta$ lines seen in our Galactic sample, have statistical significances $\ge 3\sigma$ (SAX~J1808.4-3658, 
4U~1728-16 and V*V~821 Ara), and for only 2 of these we can also study the associated OII K$\alpha$ transition: 4U~1728–16 (Fig. 
\ref{galspec}, top panel) and V*V~821 Ara (Fig \ref{galspec}, middle panel). 
We therefore decided to coadd the RGS spectra of the remaining 18 X-ray binaries of our sample, and evaluate the strength of the 
OI K$\alpha$, OII K$\alpha$ and OII K$\beta$ lines in this coadded spectrum (hereinafter ``18CB'' spectrum: Fig. \ref{galspec}, 
bottom panel, and last row of Table 3). 
The 18CB spectrum has SNRE=152 at 22.2 \AA, and detects all three OI and OII lines at high statistical significances: 93$\sigma$ (OI 
K$\alpha$), 70$\sigma$ (OII K$\alpha$) and 7.1$\sigma$ (OII K$\beta$: Table 3, last row). 
As column of neutral gas N$_H^X$ attenuating this virtual line of sight we assume the Signal-to-Noise weighted average of the 18 best-fitting 
N$_H^X$ values listed in Table 1 (3$^{rd}$ column): $<$N$_H^X$(18CB)$> = (2.0 \pm 0.3) \times 10^{21}$ cm$^{-2}$. 

In the following discussion we consider only these three spectra as representative of our Galactic lines of sight: the two spectra of the 
Galactic X-ray binaries 4U~1728-16 and V*V~821 Ara, plus the coadded spectrum of the remaining 18 X-ray binaries, 18CB. 

\subsection{The Coadded HETG and LETG Spectra of the AGNs}
Contrarily to the Galactic case, for our extragalactic targets the EWs of the K$\alpha$ and K$\beta$ transitions of OII are generally 
(with the exception of our two reference targets) comparable (Fig. \ref{ewoikaoiikaoiikbvsnh}, middle and bottom panels), and so are the 
associated statistical significances of these detections. 
This may indicate a higher level of saturation (see \S 3.4) of these lines, but may also possibly be a consequence of the generally lower SNRE of 
the G-sample spectra, compared to both the Galactic ones and those of our 2 reference extragalactic targets, which introduces the 
spectral analogous of the ``Eddington bias'' in the measurements of line EWs when hinted in the data (Senatore et al., in preparation). 
Indeed only along two of the 29 lines of sight to the extragalactic sources of our sample, the OII K$\beta$ is detected at statistical significance 
$\ge 3\sigma$ (the lines of sight toward Mkn~421 and Mkn~501), but only one of these two spectra (Mkn~421) can also detect the associated 
OII K$\alpha$. 
The situation is only slightly better when we consider the combined (in quadrature) statistical significance of the associated OII lines 
(the K$\alpha$ and the K$\beta$). 
In this case, the spectra of our two reference targets detect the OII absorbers at combined statistical significances 
of 17.4$\sigma$ (Mkn~421) and 8$\sigma$ (PKS~2155-304), but the remaining 5 OII K$\alpha$ and K$\beta$ associations have combined 
statistical significances of 3.1$\sigma$  (H~2356-309), 2.9$\sigma$ (1ES~0120+340), 3.9$\sigma$ (3C~454.3) 2.7$\sigma$ (H~2106-099) 
and 2.1$\sigma$ (H~1821+643), all well below the threshold of $4.2\sigma$, which corresponds to the combined statistical significance of 
two lines detected at 3$\sigma$. 
We therefore decided to coadd all the spectra of the G-sample taken with the same spectrometer (LETG and HETG) and that do not 
show contamination by intrinsic warm absorber/emitter components at the relevant wavelengths of our OI and OII transitions, with the 
exception of those of our two reference targets, and evaluate the strength of the OI K$\alpha$, OII K$\alpha$ and OII K$\beta$ lines in these 
two higher SNRE spectra (last 2 rows of Table 4 and bottom panel of Fig. \ref{exgalspec}). 
The Coadded LETG spectrum (hereinafter ``8CLETG'' spectrum: Fig. \ref{exgalspec}, bottom panel) is then the sum of 8 LETG spectra of the 
G-sample (all but those of MR~2251-178, Mrk~279 and NGC~4051), while the coadded HETG spectrum (hereinafter ``9CHETG'' spectrum) is the 
sum of 9 HETG spectra of the G-sample (all but those of Ark~564, Mrk~290, Mrk~509, NGC~3516, NGC~3783 and NGC~4593).  
The 8CLETG spectrum has SNRE=18 at 22.2 \AA, and detects the three OI K$\alpha$ and OII K$\alpha$ and K$\beta$ lines at relatively 
high statistical significances (4.3, 5 and 2.8$\sigma$, respectively), with a combined OII K$\alpha$ and K$\beta$ detection at a 
5.7$\sigma$ significance level (Table 4, last but one row). 
On the contrary, the 9CHETG spectrum has only SNRE=5.8 at 22.2 \AA, not sufficient to clearly detect the weak OII K$\beta$ line 
that is only hinted at in the data with a 3$\sigma$ EW upper limit of $<27.7$ m\AA\ (Table 4, last row). 

In the following discussion we therefore consider only these three spectra as representative of our extragalactic lines of sight: the spectra 
of our two reference targets Mkn~421 and PKS~2155-304, plus the coadded LETG spectrum 8CLETG. 
Analogously with the Galactic case, also for the 8CLETG spectrum we assume a column of neutral gas equal to the Signal-to-Noise 
weighted average of the 8 best-fitting N$_H^X$ values listed in Table 2 (4$^{rd}$ column): $<$N$_H^X$(8CLETG)$> = (0.8 \pm 0.3) 
\times 10^{21}$ cm$^{-2}$. 

\subsection{The EW(OII K$\alpha$) / EW(OII K$\beta$) Ratio}
For two unsaturated lines of the same ion, the EW ratio is equal to the product between the oscillator strength ratio 
and the square of the rest-frame wavelength ratio. For the K$\alpha$ and K$\beta$ transitions of OII, we then expect 
EW(OII K$\alpha$) / EW(OII K$\beta$) = (f$_{OII K\alpha}$ / f$_{OII K\beta}$) [$\lambda$(OII K$\alpha$) / $\lambda$(OII K$\beta$)]$^2$ $\simeq 
8.4$, where we used $f_{OII K\alpha} = 0.2$, $f_{OII K\beta} = 0.026$ (Behar, private communication). 

\begin{figure} 
\begin{center}
\includegraphics[width=84mm]{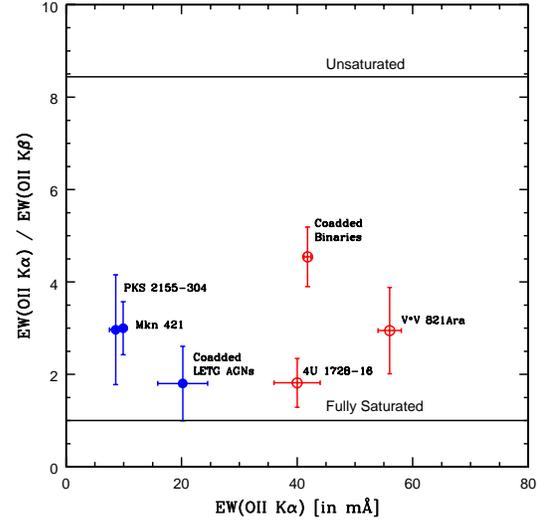}
\end{center}
\caption{ Measured EW(OII K$\alpha$) / EW(OII K$\beta$) ratio for the 6 high S/N spectra of our Galactic and extragalactic sample, as a 
function of the measured EW(OII K$\alpha$). See text for details.}
\label{ewkaewkbratio}
\end{figure}

Figure \ref{ewkaewkbratio} shows the measured EW(OII K$\alpha$) / EW(OII K$\beta$) ratio for the Galactic (red open circles) and extragalactic (blue solid circles) 
lines of sight (or averaged lines of sight for the coadded spectra 18CB and 8CLETG), as a function of the measured EW(OII K$\alpha$).  
The two horizontal lines in Fig. \ref{ewkaewkbratio} mark the limits of full saturation (EW(OII K$\alpha$) / EW(OII K$\beta$) = 1) and 
no saturation (EW(OII K$\alpha$) / EW(OII K$\beta$) = 8.4). 
The EW(OII K$\alpha$) / EW(OII K$\beta$) ratio settles around consistent averages of $3.1 \pm 1.4$ and $2.6 \pm 0.6$ for the Galactic and extragalactic 
lines of sight, respectively, but correponds to OII K$\alpha$ EWs that are systematically higher for the Galactic ($<$EW(OII K$\alpha$)$>_{gal} = 46 \pm 8$) than 
the extragalactic ($<$EW(OII K$\alpha$)$>_{exgal} = 13 \pm 6$) sightlines. 
This requires that either the average line-of-sight velocity dispersion or column density (or both) of the OII absorber, be intrinsically 
different in the two samples. 

\subsection{Column Density and Doppler Parameter of the OII Absorber in the two Samples}
The six spectra of our Galactic and extragalactic samples have sufficient statistics in their associated OII K$\alpha$ and K$\beta$ lines 
to perform a detailed curve of growth analysis on their EW ratios and break the implicit degeneracy between the ion column density 
and the Doppler parameter of the absorber along these lines sight (or ``S/N-averaged'' lines of sight in the case of the two co-added Galactic 
and extragalactic spectra, 18CB and 8CLETG). 

For each of the 6 available lines of sight we run curve-of-growth analyses for the K$\alpha$ and K$\beta$ transitions of OII, over a grid of Doppler parameter 
values comprised between $b=0-300$ km s$^{-1}$, with a resolution of 10 km s$^{-1}$. For each value of the Doppler parameter within the grid, we evaluate the 
column density of OII corresponding to the best-fitting EW value of the K$\alpha$ and K$\beta$ absorption lines, plus and minus their 1$\sigma$ statistical 
uncertainties. 
This generates, for each line of sight, two regions of $\pm 1 \sigma$ allowed intervals for each of the two transitions, in the N$_{OII}$-$b$ plane, and the 
intersection of these two regions defines the $\pm 1\sigma$ N$_{OII}$-$b$ solution interval
\footnote{We have previously exploited this technique succesfully in Williams et al. (2005, 2007) and Gupta et al. (2012)}.
. 

\begin{figure*} 
\begin{minipage}{126mm}
\begin{center}
\includegraphics[width=120mm]{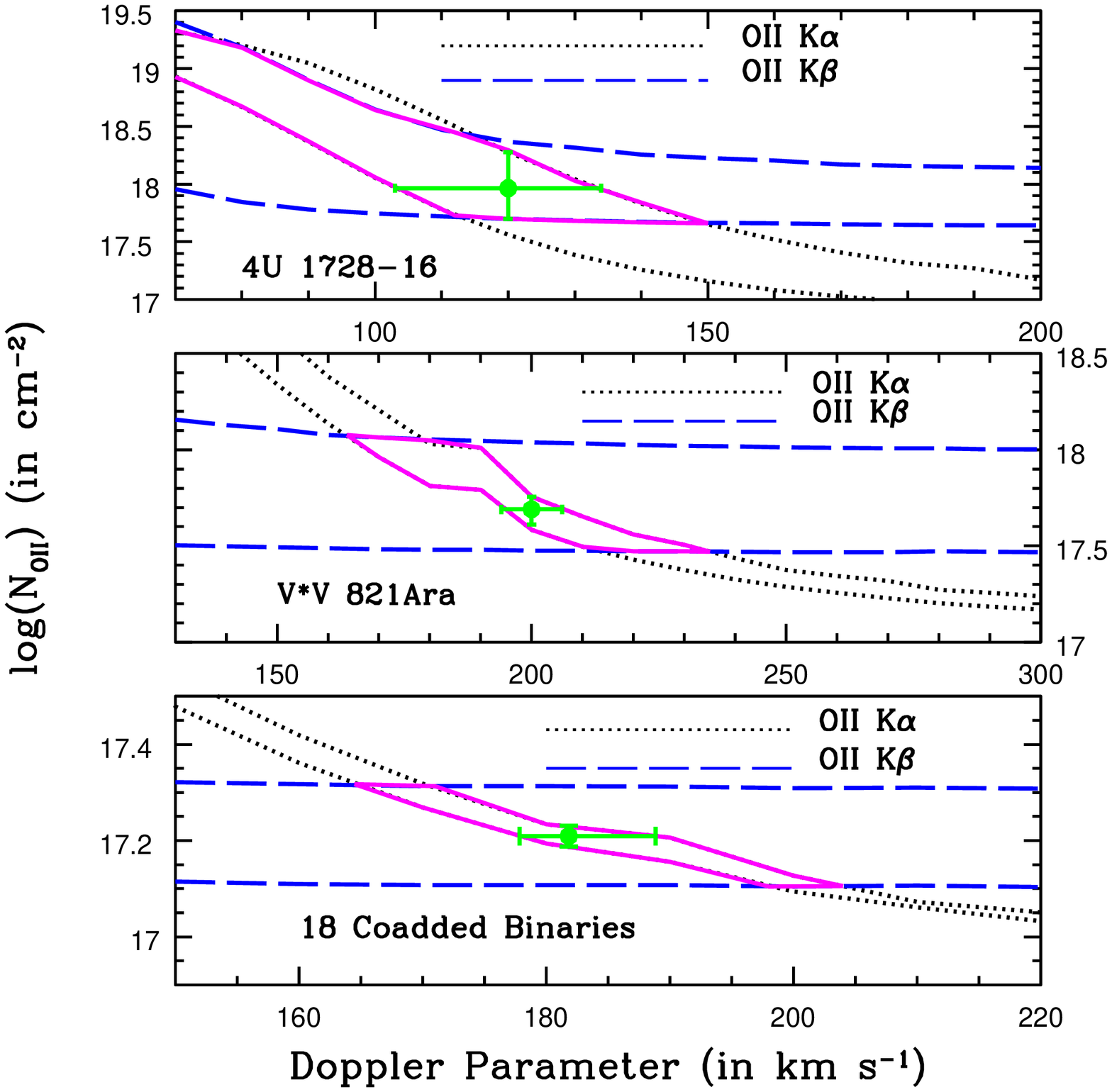}
\end{center}
\caption{ $\pm 1 \sigma$ allowed regions for the K$\alpha$ (dotted black) and K$\beta$ (dashed blue) transitions of OII, in the N$_{OII}$-$b$ plane, for 
the lines of sight towards our Galactic targets. See text for details.}
\label{Galacticnvsb}
\end{minipage}
\end{figure*}

\begin{figure*} 
\begin{minipage}{126mm}
\begin{center}
\includegraphics[width=120mm]{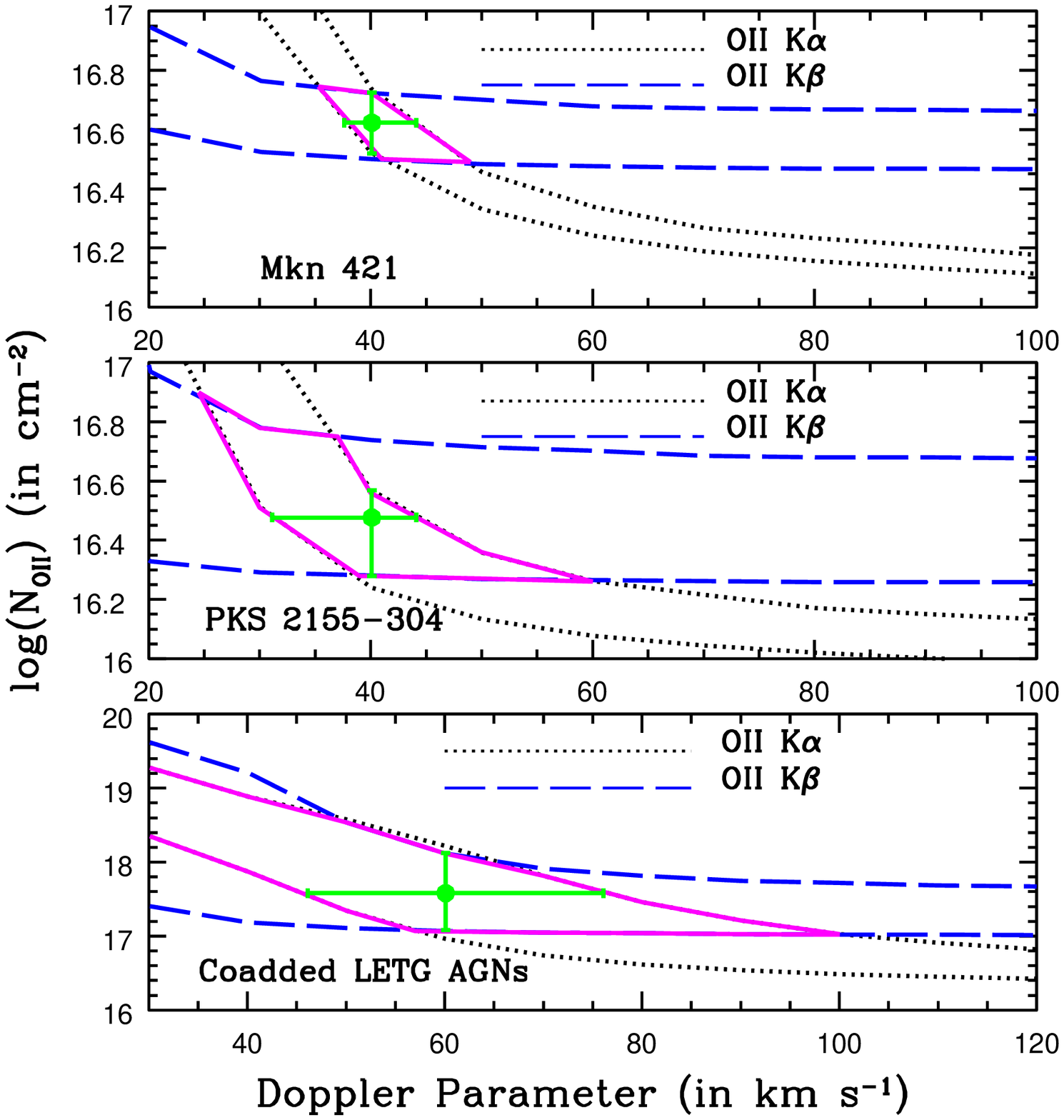}
\end{center}
\caption{ $\pm 1 \sigma$ allowed regions for the K$\alpha$ (dotted black) and K$\beta$ (dasged blue) transitions of OII, in the N$_{OII}$-$b$ plane, for 
the lines of sight towards our extragalactic targets. See text for details.}
\label{Extragalacticnvsb}
\end{minipage}
\end{figure*}

Figures \ref{Galacticnvsb} and \ref{Extragalacticnvsb} show the result of this procedures for our 3 Galactic and 3 extragalactic lines of sight (or S/N-weighted 
average lines of sight for the 18CB and 8CLETG spectra), respectively, while Table 5 summarizes the N$_{OII}$ and $b$ measurements.  
In each figure, the dotted black and dashed blue curves delimit the $\pm 1\sigma$ allowed K$\alpha$ and K$\beta$ regions, respectively, while their 
intersection (solid magenta contour) defines the allowed 1$\sigma$ N$_{OII}-b$ contour and solution. 
In each figure, the green point marks the solution corresponding to the curve of growths computed for the best fitting EWs of the two transitions. 

The gas producing the OII absorption along the Galactic lines of sight (i.e. within the Galactic disk) is characterized by OII column densities, as 
well as Doppler parameters, that are systematically higher than those of the gas imprinting OII absorption in the direction of the extra-galactic line of sight. 
The average Doppler parameters measured against the Galactic and extragalactic targets are $<b_{gal}> = 170 \pm 40$ km s$^{-1}$ and $<b_{exgal}> = 50 \pm 10$, 
respectively. $<b_{gal}>$ is of the order of the rotational velocities in the disk of the Galaxy and thus the OII absorber detected along the Galactic lines of sight, 
is plausibly associated with interstellar matter filling the Galaxy's disk. 
The lower Doppler velocity measured against the extra-galactic targets, as well as the smaller OII columns, could simply reflect the fact that our 
extragalactic targets are all located at high Galactic latitudes and so cross a much lower portion of the Galactic disk. However, these differences could also 
in principle indicate that the bulk of the absorption produced along the extraglactic lines of sight has a different origin than that produced in the disk of 
the Galaxy, against our Galactic targets.  

\begin{table} 
\footnotesize
\begin{center}
\caption{\bf OII Column Densities and Doppler Parameters}
\vspace{0.4truecm}
\begin{tabular}{|c|cc|}
\hline
Line-of-Sight & N$_{OII}$ & b \\
& in 10$^{17}$ cm$^{-2}$ & in km s$^{-1}$ \\ 
\hline
\multicolumn{3}{|c|} {Galactic Lines of Sight} \\
\hline 
4U~1728-16 & $9^{+10}_{-4}$ & $120^{+14}_{-17}$ \\
V*V~821 Ara & $4.9 \pm 0.8$ & $200 \pm 6$ \\
18CB & $1.62 \pm 0.08$ & $182^{+7}_{-4}$ \\
\hline
\hline
\multicolumn{3}{|c|} {Extragalactic Lines of Sight} \\
\hline 
Mkn~421 & $0.42^{+0.11}_{-0.09}$ & $40^{+4}_{-3}$ \\
PKS~2155-304 & $0.30^{+0.07}_{-0.11}$ & $40^{+4}_{-9}$ \\
8CLETG & $4^{+9}_{-3}$ & $60^{+16}_{-15}$ \\
\hline
\end{tabular}
\end{center}
\end{table}

\subsection{Self-Consistent Physical Modeling of the OI-OII Absorbers}
To discriminate between these two hypotheses, the physical properties of the two potentially distinct absorbers need to be derived. 
This requires modeling of both the OI and OII absorbers in a self-consistent manner and in the framework of ISM/CGM physics (i.e. photo-ionization by Galactic 
and extragalactic diffuse radiation fields), that is with a model that includes accurate treatments of both (a) the ISM/CGM photo-ionization structure and (b) 
line saturation (for all the main bound-bound transitions from neutral and ionized species), i.e. the evaluation of the exact 
line profile as a function of its column density and intrinsic Doppler parameter. 

There are no publicly available spectral models of the ISM that include both these ingredients. 
The {\em absline} model introduced by Yao \& Wang (2005) as a single-line Voigt profile routine, and then adapted by Yao \& Wang (2009) to a more general case of a 
multi-line routine modeling simultaneously a number of transitions from different ions linked to each other through physically-motivated relative ion fractions, is optimized to the Galaxy's ISM but is not publicly available. 
The new high resolution version {\em tbnew}
\footnote{http://pulsar.sternwarte.uni-erlangen.de/wilms/research/tbabs/}
of the XSPEC-native {\em tbabs} model (Wilms et al., 2000), improves the spectral resolution of the cross-sections used in the code, but contains only 
cross-sections from neutral ions and does not compute line-saturation (it simply multiplies the cross section by the column of absorbing gas to produce 
the effective optical depth at the center of the line and convolves that, in the fit, with the Instument Line Spread Function: LSF). 
On the other hand, the recently published {\em ISMabs} model (Gatuzz et al., 2014), contains both neutral and ionized species but fits them independently 
(not within a defined gas ionization structure) and does not treat line saturation either. 
Finally, the SPEX code (Kaastra, Mewe \& Nieuwenhuijzen (1996) contains several absorption models none of which, however, is optimized for cold to mild 
ionized ISM/CGM components. 
For example, the model {\em hot} can be adapted to study the cold and neutral phase of the ISM by freezing the tempaerature of this model to its 
lowest avaialable value (e.g. Pinto et al., 2013), but it is collisoinally ionized. The {\em slab} model, like {\em ISMabs}, fits each ion independently and not with a 
(physicall motivated gas ionization structure. Last, {\em Xsabs} is a photo-ionization model that is optimized by default to a typical Seyfert-1 photo-ionizing 
SED, and thought it could be in principle be adapted to a different photo-ionizing SED, does not allow the mixing of two diverse photo-ionizing radiation fields 
as in the case of disk and halo ISM/CGM.

We therefore decided to adapt our  PHotoionized Absorber Spectral Engine (PHASE, Krongold et al., 2003) code, originally developed for the study of highly ionized 
Warm Asborbers in Active Galactic Nuclei, to our ISM case-study (hereinafter {\em galabs} model). 
We did this by running our code using as photoionizing source a combination of the external meta-galactic ionizing background radiation at $z=0$ (e.g. 
Cen \& Fang, 2006: see their Figure 1, magenta curve) and the unextinguished FUV-X-ray Galactic foreground emission from the disk (as tabulated in 
{\em Cloudy} - Ferland et al., 2013 - in the ``table ISM'' command: see Figure 5 in the {\em Cloudy} ``Hazy 1'' manual), and computed a grid of 
photoionization models for 46 values of the ionization parameter (defined as the ratio of the volume density of $E\ge 13.6$ eV photons to the gas volume 
density) and 51 values of the equivalent H column density. 
Within the disk of the Galaxy, and at 5 kpc from the Galactic center (see, e.g. Mezger, 1990), the FUV Galactic diffuse emission (mostly provided by O and B stars 
in the Galaxy's plane: e.g. Haffner et al., 2009, and references therein) largely dominates (by $\sim 2$ orders of magnitude) over the extragalactic contribution at $\lambda = 1000$ \AA, 
but at shorter wavelengths the hot star contribution decreases more steeply than the AGN contribution in the extragalactic radiation field, and in the soft X-rays the two radiation fields 
contribute rougly equally. In the Galaxy halo, at large Galactocentric distances, the extragalactic contribution dominates the ionizing source of radiation. 
Since the volume densities of ionizing photons in the disk and halo of our Galaxy are fixed (i.e. the normalizations of the ionizing radiation), the value of the 
ionization parameter defines the volume density of the gas and it is uniquely associated to a photoionization equilibrium temperature of the gas. 

Our {\em galabs} model uses the most up-to-date bound-bound wavelengths and cross-sections for the inner-shell K transitions of neutral and ionized 
light metals (e.g. Juette et al., 2004, Gatuzz et al., 2013a,b, Kallman, 2015, in XSTAR v2.2
\footnote{ftp://legacy.gsfc.nasa.gov/software/plasma\_codes/xstar/rates/\\xstarlines.text}
, Behar, private communication), but, compared to {\em tbabs}, adopts a poorer description of the atomic and molecular neutral bound-free cross-sections. 
We therefore use {\em galabs} only to fit the bound-bound resonant transitions detected in the 6 spectra of our samples (together with those not explicitly detected 
but present in the {\em galabs} model, like the OI K$\beta$ and K$\gamma$, or the OIII K$\alpha$ triplet, but at whose wavelengths the data provide stringent upper 
limits), and combine it with {\em tbabs} to model the neutral bound-free transitions. In both models we adopt the revised Wilms et al. (2000) ISM abundances 
(e.g. $A_O = 4.9 \times 10^{-4}$ for oxygen). 

\subsection{Physics, Chemistry and Mass of the Absorbers}
Since both OI and OII are detected towards both our Galactic and extragalactic targets, at least part of the material imprinting such features must 
be mildly ionized. The ionization degree, and thus the temperature, of the OI-OII absorbers is unequivocally determined by the fractional ratio of the two ions, 
$f_{OII}/f_{OI}$. This fraction is simply given by the ion column density ratio: $f_{OII}/f_{OI} =$N$_{OII}/$N$_{OI}$. 

\subsubsection{Model A: Fit with a Single {\em galabs} component}
\begin{figure} 
\begin{center}
\includegraphics[width=84mm]{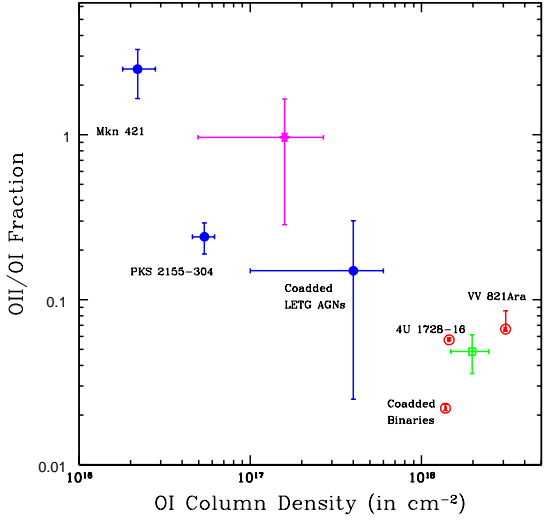}
\end{center}
\caption{ $f_{OII}/f_{OI}$ fraction for the three Galactic (red open circles) and three extragalalactic (blue solid circels) lines of sight of our samples, as a function 
of the OII column density measured towards these directions. The starred magenta point and the green square are the averages Galactic and extragalactic 
line-of-sight values.}
\label{noiinoifracvsnoi}
\end{figure}

To derive the columns of OI and OII along our 6 Galactic and extra-galactic lines of sight, we first fitted our spectra with a single {\em galabs} 
component, multiplied by {\em tbabs}: Model A. For each line of sight, we linked the equivalent H column densities of {\em galabs} and {\em tbabs} to the 
same value (which enables metallicity estimates), while allowing the oxygen abundance to freely vary below the Solar value (i.e. we do not allow for super-Solar 
metallicities). 
In Table 6, we list, for each of the 6 lines of sight, the best-fitting OI and OII column densities and Doppler parameters together with the derived OII/OI 
ionization fractions, the corresponding best-fitting temperatures and the metallicities of the absorbers. 
\begin{table*} 
\footnotesize
\begin{center}
\caption{\bf Model A: Physical Properties of the OI-OII Absorbers}
\vspace{0.4truecm}
\begin{tabular}{|c|cccccc|}
\hline
Line-of-Sight & N$_{OI}$ & N$_{OII}$ & b & $f_{OII}/f_{OI}$ & T$_{phot}$ & $Z$ \\
& in 10$^{17}$ cm$^{-2}$ & in 10$^{17}$ cm$^{-2}$ & in km s$^{-1}$ & & in K & in $Z_{\odot}$ \\ 
\hline
\multicolumn{7}{|c|} {Galactic Lines of Sight} \\
\hline 
4U~1728-16 & $14.5^{+0.1}_{-0.2}$ & $0.829^{+0.018}_{-0.001}$ & $213 \pm 14$ & $0.0572^{+0.0014}_{-0.0004}$ & $600^{+180}_{-90}$ & $0.79 \pm 0.06$ \\
V*V~821 Ara & $31.0^{+0.1}_{-0.6}$ & $2.06^{+0.60}_{-0.05}$ & $268 \pm 11$ & $0.066^{+0.019}_{-0.002}$ & $900^{+300}_{-130}$ & $0.87 \pm 0.02$ \\
18CB & $13.80 \pm 0.01$ & $0.30^{+0.02}_{-0.01}$ & $186 \pm 2$ & $0.0220^{+0.0013}_{-0.0007}$ & $310 \pm 10$ & $0.8 \pm 0.1$ \\
\hline
\hline
\multicolumn{7}{|c|} {Extragalactic Lines of Sight} \\
\hline 
Mkn~421 & $0.22^{+0.6}_{-0.4}$ & $0.55^{+0.14}_{-0.11}$ & $39 \pm 1$ & $2.5 \pm 0.8$ & $4700 \pm 200$ & $0.12^{+0.04}_{-0.02}$ \\
PKS~2155-304 & $0.54 \pm 0.08$ & $0.13 \pm 0.02$ & $45 \pm 2$ & $0.24 \pm 0.05$ & $1700 \pm 300$ & $0.39 \pm 0.06$ \\
8CLETG & $4^{+2}_{-3}$ & $0.6 \pm 0.4$ & $60 \pm 13$ & $0.2^{+0.2}_{-0.1}$ & $2300^{+2000}_{-1300}$ & $0.6^{+1.8}_{-0.5}$ \\
\hline
\end{tabular}
\end{center}
\end{table*}
The measured $f_{OII}/f_{OI}$ fractions are plotted in Fig. \ref{noiinoifracvsnoi}, together with their average Galactic (starred magenta point) and 
extragalactic (green open square) line-of-sight values, against the best-fitting OI columns. 
The OI-OII absorber probed by the extragalactic lines of sight has average $<f_{OII}/f_{OI}>_{Exgal} = 1.0 \pm 0.7$, $<$N$_{OI}^{Exgal}> = (2 \pm 1) \times 
10^{17}$ cm$^{-2}$ and $<$N$_{OII}^{Exgal}> = (4 \pm 1) \times 10^{16}$ cm$^{-2}$. 
The same quantities for the Galactic lines of sight are: $<f_{OII}/f_{OI}>_{Gal} = 0.05 \pm 0.01$, $<$N$_{OI}^{Gal}> = (2.0 \pm 0.5) \times 
10^{18}$ cm$^{-2}$ and $<$N$_{OII}^{Gal}> = (1.1 \pm 0.5) \times 10^{17}$ cm$^{-2}$. 
Thus, while $<$N$_{OII}^{Exgal}>$ and $<$N$_{OII}^{Gal}>$ are consistent with each other within 1.2$\sigma$ (despite the much larger portion of Galactic disk 
crossed by the Galactic lines of sight compared to the extragalactic ones: \S 3.7.3 and Table 9),  both the average ionization degree and OI column density are 
neatly separated for Galactic and extragalactci targets, indicating that the extragalactic line-of-sight absorber is on average more ionized than that observed along 
the Galactic lines of sight. 

\begin{figure} 
\begin{center}
\includegraphics[width=84mm]{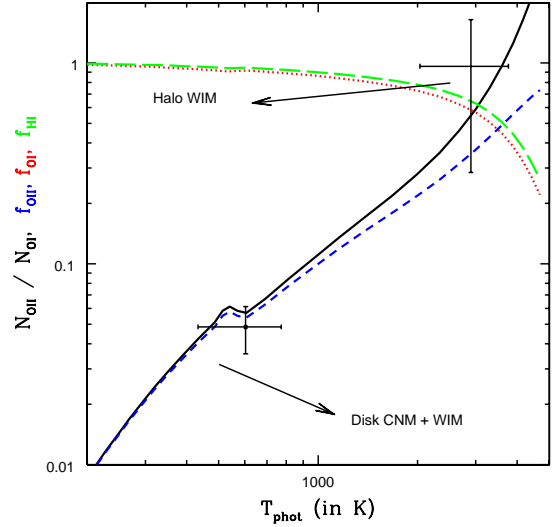}
\end{center}
\caption{Measured $f_{OII}/f_{OI}$ ion fractions superimposed on the predicted relative fractions of HI (green, long-dashed), OI (red, dotted), OII (blue, 
short-dashed), as well as the $f_{OII}/f_{OI}$ ratio (solid black), and plotted against the photoionization temperature.}
\label{noiinoifracvst}
\end{figure}

\noindent 
This can be easily seen in Fig. \ref{noiinoifracvst}, where the average values of $f_{OII}/f_{OI}$ for the Galactic and extragalactic lines of sight (black points and 
errorbars) are plotted as a function of the respective average ionization temperatures T$_{phot}$ ($<$T$_{phot}^{Gal}> = 600 \pm 200$ K, $<$T$_{phot}^{Exgal}> = 
2900 \pm 900$ K, respectively) superimposed on the $f_{OII}/f_{OI}$ ratio (solid black) (as well as the relative fractions of the HI - long-dashed green -, OI - 
dotted red - and OII - short-dashed blue - ions) predicted by our {\em galabs} model. 

\subsubsection{The Need for an Extra {\em galabs} Component for the Galactic Lines of Sight: Model B}
A comparison between the OII column densities inferred from our best-fitting models A (Table 6) and those derived through curve of growth analysis of 
the best fittting Gaussian OII K$\alpha$ and K$\beta$ EWs (Table 5, \S 3.5), already suggests that Model A is not an accurate description of the 
OI-OII absorber detected towards our Galactic targets. Indeed, while the N$_{OII}$ measured with the two methods towards the extragalactic targets are 
all similar and consistent with each others within at most 1.5$\sigma$ (for PKS~2155-304), those measured with Model A against the Galactic targets are 
all systematically lower than the same quantities measured through single-Gaussian fitting.
These differences are confirmed by a comparison of the of the best-fitting OI K$\alpha$ and OII K$\alpha$ and K$\beta$ EWs obtained with our Model A, and 
the same quantities measured by fitting independent absorption Gaussians to our data (Table 7). 
\begin{table*} 
\footnotesize
\begin{center}
\caption{\bf Comparison of OI K$\alpha$, OII K$\alpha$ and OII K$\beta$ EWs obtained with Gaussians and with Model A}
\vspace{0.4truecm}
\begin{tabular}{|c|ccc|ccc|}
\hline
Line-of-Sight & EW$_{OI K\alpha}^{Model A}$ & EW$_{OII K\alpha}^{Model A}$ & EW$_{OII K\beta}^{Model A}$  & EW$_{OI K\alpha}^{Gauss}$ & EW$_{OII K\alpha}^{Gauss}$ & 
EW$_{OII K\beta}^{Gauss}$ \\
& in m\AA\ & in m\AA\ & in m\AA\ & in m\AA\ & in m\AA\ & in m\AA\ \\ 
\hline
\multicolumn{7}{|c|} {Galactic Lines of Sight} \\
\hline 
4U~1728-16 & $67 \pm 4$ & $39^{+5}_{-1}$ & $8.5^{+1.6}_{-0.1}$ & $76 \pm 5$ & $40 \pm 4$ & $22 \pm 6$\\
V*V~821 Ara & $91 \pm 3$ & $62.6^{+0.1}_{-2.7}$ & $18.5^{+4.5}_{-0.5}$ & $110 \pm 5$ & $56 \pm 2$ & $19 \pm 6$ \\
18CB & $59.3^{+0.9}_{-0.2}$ & $20.8^{+1.0}_{-0.7}$ & $3.3^{+0.2}_{-0.1}$ & $74.4 \pm 0.8$ & $41.8 \pm 0.6$ & $9.2 \pm 0.3$ \\
\hline
\hline
\multicolumn{7}{|c|} {Extragalactic Lines of Sight} \\
\hline 
Mkn~421 & $6.6^{+0.8}_{-0.3}$ & $13 \pm 1$ & $4.3^{+0.6}_{-0.3}$ & $8.5 \pm 0.5$ & $9.9 \pm 0.6$ & $3.3 \pm 0.6$\\
PKS~2155-304 & $10.5^{+1.2}_{-0.4}$ & $7.8^{+0.8}_{-0.1}$ & $1.40^{+0.20}_{-0.01}$ & $9.8 \pm 1.0$ & $8.6 \pm 1.1$ & $2.9 \pm 1.1$ \\
8CLETG & $20^{+11}_{-8}$ & $17^{+8}_{-4}$ & $5^{+5}_{-1}$ & $17 \pm 4$ & $20 \pm 4$ & $11 \pm 4$ \\
\hline
\end{tabular}
\end{center}
\end{table*}

\begin{figure} 
\begin{center}
\includegraphics[width=84mm]{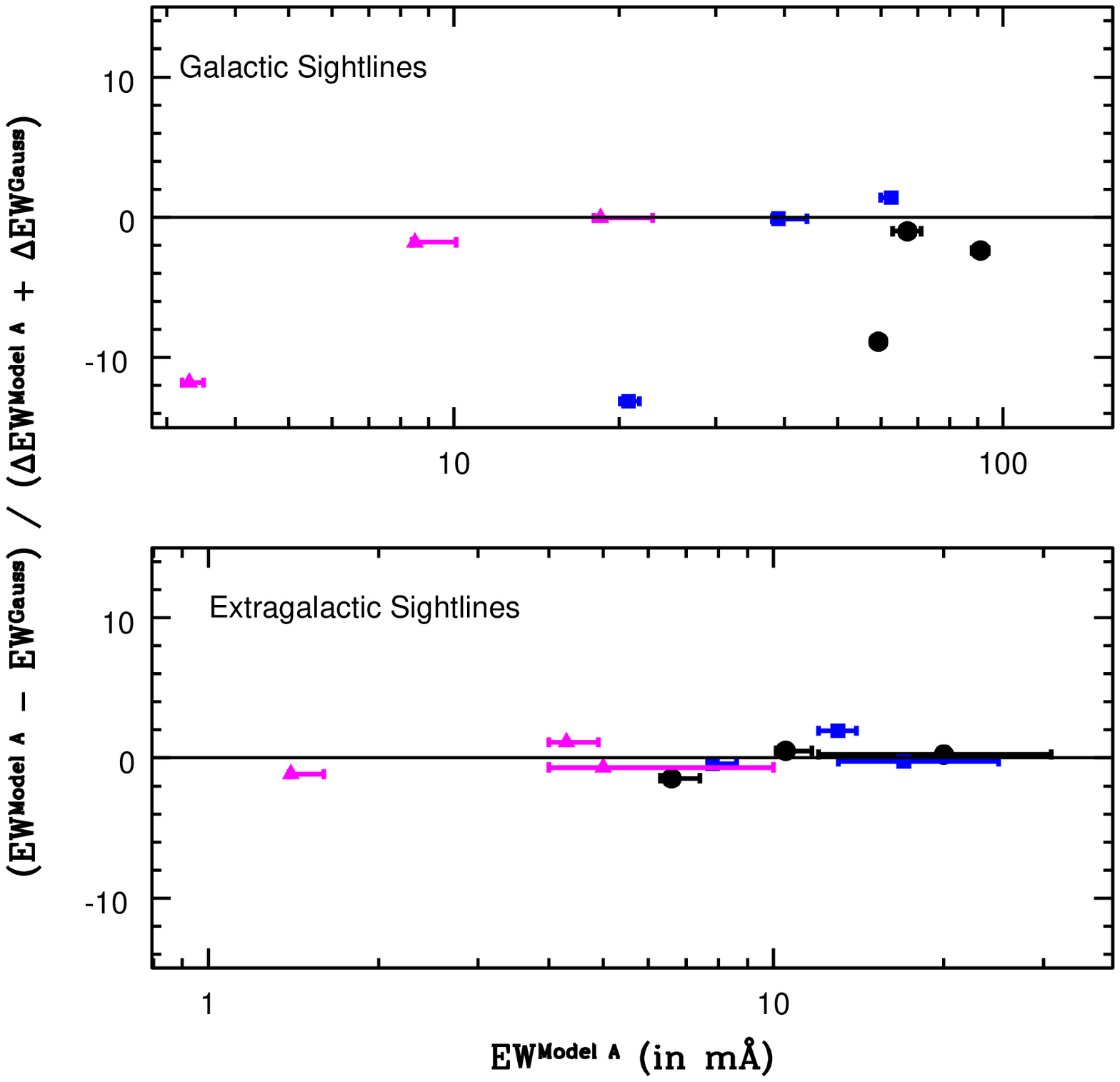}
\end{center}
\caption{Differences, in units of standard deviations, between the OI K$\alpha$ (black circles), OII K$\alpha$ (blue squares) and OII K$\beta$ (magenta triangles) 
EWs measured with Model A and those measures with independent single-line Gaussians, for Galactic (top panel) and extragalactic (bottom panels) sightlines.}
\label{ewdiff_sigma}
\end{figure}

The same comparison is visually shown in Fig. \ref{ewdiff_sigma}, where the differences between the EWs measured with Model A and those measures with 
independent single-line Gaussians is plotted, in units of standard deviations, against the Model-A EWs, for the three oxygen lines, OI K$\alpha$ (black circles), 
OII K$\alpha$ (blue squares) and OII K$\beta$ (magenta triangles), and for Galactic (top panel) and extragalactic (bottom panels) sightlines. 
Clearly, while for the extragalactic targets all Model-A EW measurements deviate $<2$$\sigma$ from the corresponding Gaussian EW estimates, for the Galactic 
targets, the majority (5 out of 9) of the Model-A EW measurements deviate $>2\sigma$ from their corresponding Gaussian estimates. 

This, on one hand indicates that the bulk of the OI-OII absorption seen in the spectra of our extragalactic targets is indeed imprinted by mildly ionized gas 
(hereinafter Warm Ionized Metal Medium, or WIMM) with an average temperature of $<T_{Exgal}>^{WIMM} = 2900 \pm 900$ K , an average metallicity of 
$<Z_{Exgal}>^{WIMM} = (0.4 \pm 0.1) Z_{\odot}$ and located outside (above and below) the Galactic disk, and on the other suggests that the Galaxy disk (where our 
Galactic targets seat) is permeated by at least two distinct OI-OII absorbers which, if modeled with a single gaseous component, are approximated by a 
quasi-neutral medium with average temperature and metallicity $<$T$_{Gal}>^{1-Component} = 600 \pm 200$ K and $<Z_{Gal}>^{1-Component} = (0.82 \pm 0.02) 
Z_{\odot}$. 

\begin{figure} 
\begin{center}
\includegraphics[width=84mm]{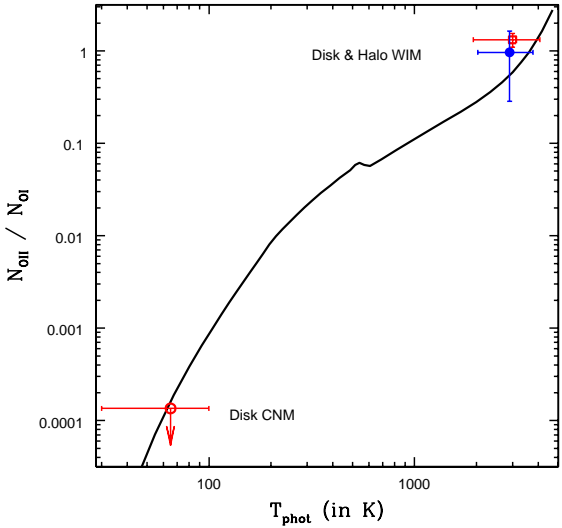}
\end{center}
\caption{Measured $f_{OII}/f_{OI}$ ion fractions for the three CNMM (red open circle), Disk WIMM (red open square) and Halo WIMM (solid blue circles) solutions, 
superimposed on the predicted $f_{OII}/f_{OI}$ ratio (black) and plotted against their photoionization temperature}
\label{threecompsol}
\end{figure}

To test this hypothesis we added an additional {\em galabs} component to our best-fitting Model A and re-fitted our data leaving all parameters free to vary 
independenly, except the sum of the two {\em galabs} equivalent H column densities that is linked to the {\em tbabs} equivalent H column density 
N$_H^X$ (which, again, enables metallicity estimates; Model B). 
For all three extragalactic lines of sight, the additional {\em galabs} is not statistically required (i.e. the best fitting equivalent H column density is zero). 
For our Galactic lines of sight, instead, 2-components solutions are found, and the temperature of one of the two components is systematically 
pegged to its minimum tabulated value in our {\em galabs} model, i.e. 30 K. 
To reduce the statistical errors on the measured ion column densities of our best fitting Model B for our Galactic lines of sight, we therefore forced the 
temperature of one of the two components to vary only between 30-100 K, that is to be neutral (HI and OI fractions are $>0.9999$), and refitted our XRB 
spectra. The best fitting Model B parameters of our Galactic lines of sight are summarized in Table 8. 
Figure \ref{threecompsol} also shows the three average $f_{OII}/f_{OI}$ ionization fraction and photoionization temperature of the these three solutions (the 
ionized solutions along the extragalactic lines of sight - \S 3.7.1, Table 6 - and the cold and ionized solutions along the Galactic lines of sight - Table 8). 
\begin{table*} 
\footnotesize
\begin{center}
\caption{\bf Model B: Physical Properties of the OI-OII Absorbers along Galactic Lines of Sight}
\vspace{0.4truecm}
\begin{tabular}{|c|cccccc|}
\hline
Line-of-Sight & N$_{OI}$ & N$_{OII}$ & b & $f_{OII}/f_{OI}$ & T$_{phot}$ & Metallicity \\
& in 10$^{17}$ cm$^{-2}$ & in 10$^{17}$ cm$^{-2}$ & in km s$^{-1}$ & & in K & in Solar units \\ 
\hline
\multicolumn{7}{|c|} {Galactic Lines of Sight: Cold Neutral Metal Medium (CNMM) Component} \\
\hline 
4U~1728-16 & $4.46 \pm 0.03$ & $<10^{-4}$ & $265_{-23}^{+17}$ & $<10^{-4}$ & $30-100$ & $1.4 \pm 0.5$ \\
V*V~821 Ara & $48.4 \pm 0.3$ & $<10^{-4}$ & $280 \pm 5$ & $<10^{-5}$ & $30-100$ & $2.5^{+0.5}_{-0.8}$ \\
18CB & $7.870 \pm 0.005$ & $<10^{-4}$ & $200 \pm 5$ & $<10^{-4}$ & $30-100$ & $1.0 \pm 0.1$ \\
\hline
\hline
\multicolumn{7}{|c|} {Galactic Lines of Sight: Warm Ionized Medium (WIMM) Component} \\
\hline 
4U~1728-16 & $6.64 \pm 0.02$ & $4.24 \pm 0.01$ & $78 \pm 17$ & $0.639 \pm 0.002$ & $3300 \pm 600$ & $1.06^{+0.09}_{-0.11}$ \\
V*V~821 Ara & $4.58 \pm 0.07$ & $9.4 \pm 0.1$ & $127 \pm 10$ & $2.05 \pm 0.04$ & $4700 \pm 100$ & $0.57 \pm 0.03$ \\
18CB & $5.350 \pm 0.004$ & $6.790 \pm 0.005$ & $91 \pm 6$ & $0.1269 \pm 0.0001$ & $1000 \pm 100$ & $0.8 \pm 0.1$ \\
\hline
\end{tabular}
\end{center}
\end{table*}

\subsubsection{The Warm Ionized Medium (WIMM) in the Galaxy Disk and Halo}
About one third of the OI absorption seen along the lines of sight to our Galactic targets is imprinted by an Ionized component that also produces the totality of 
the OII absorption seen in these spectra. 
The ionized compoment has an average temperature $<$T$_{Gal}>^{WIMM} = 3000 \pm 1000$ K fully consistent with that of the only component observed along the 
extragalactic lines of sight ($<$T$_{Exgal}>^{WIMM} = 2900 \pm 900$ K), and only a facotr of 2 higher metallicity ($<Z_{Gal}>^{WIMM} = (0.8 \pm 0.1) Z_{\odot}$ 
versus $<Z_{Exgal}>^{WIMM} = (0.4 \pm 0.1) Z_{\odot}$). 
It could therefore be speculated that the WIMM is confined in the disk of the Galaxy, and absorbs both the Galactic and extragalactic lines of sight, with 
relative strengths that depend on the relative portions of disk crossed by the Galactic and extragalactic lines of sight, respectively. 
To test this hypothesis we need to assume a geometry. For the disk of the Galaxy we assume a geometrically thin cylinder, centered on the Galaxy 
center, with radius $R_G = 15$ kpc and half-height $h=0.5$ kpc. The halo, instead is here sketched as a sphere of $R_G = 15$ kpc radius, centered on the Galaxy. 
We further assume that our position in the Galaxy is $R_{\sun} = 8$ kpc away from the Galaxy's center along the $l=0$ direction, in the 
Galactic plane ($h=0$). Each line of sight, either Galactic or extragalactic, crosses pathlengths within the Galaxy's disk ($d_{disk}$) and halo 
($d_{halo}$), that depend, as seen by us, on its Galactic coordinates $(l,b)$, and for the lines of sight to the Galactic sources on whether the distance $D$ to 
the source is longer or shorter than the maximum disk pathlength $d_{disk}$ (in the computation of this quantity for our averaged 18CB line of sight, for those 
Galactic targets whose distance is not known, we conservatively assume the maximum Galaxy pathlength in that direction: $d_{tot} =  d_{disk} + d_{halo}$). 
In Table 9 we report the lengths $d_{disk}$ and $d_{halo}$ crossed in the direction of the lines of sight of our samples. Additionally, for our Galactic 
lines of sight we also report the distance $D$ to the background target. 
These amounts are all SNRE-weighted averages for the two co-added spectra of our samples, 18CB and 8CLETG. 
\begin{table} 
\footnotesize
\begin{center}
\caption{\bf Disk and Halo Lengths and WIMM Volume Densities}
\vspace{0.4truecm}
\begin{tabular}{|c|cccc|}
\hline
Line-of-Sight & $D$ & $d_{disk}$ & $d_{halo}$ & $n_{H}^{WIMM}$ \\
& in kpc & in kpc & in kpc & cm$^{-3}$ \\ 
\hline
\multicolumn{5}{|c|} {Galactic Lines of Sight} \\
\hline 
4U~1728-16 & 7.5 & 3.2 & 19.5 & 0.14 \\
V*V~821 Ara & 10 & 6.6 & 15.6 & 0.18 \\
18CB & 8.0 & 4.0 & 13.3 &  0.08 \\
\hline
\hline
\multicolumn{5}{|c|} {Extragalactic Lines of Sight} \\
\hline 
Mkn~421 & N/A & 0.6& 9.3 &  0.038 \\
PKS~2155-304 & N/A & 0.6 & 17.6 & 0.006 \\
8CLETG & N/A & 0.9 & 11.2 & 0.025 \\
\hline
\end{tabular}
\end{center}
\end{table}

The average portion of disk crossed by the lines of sight to the Galactic targets is $<d_{disk}>^{Gal} = 4.6$ kpc, compared with the $<d_{disk}>^{Exgal} = 0.7$ kpc, 
crossed by the lines of sight to the extragalactic targets. 
If the WIMM observed along the two types of sightlines were indeed the same, with the same particle volume density, then we 
would expect Galactic OI and OII columns on average 7 times larger than those observed toward extragalactci targets. This is not the case. 
From the estimated $d_{disk}$ and $d_{tot}$ we can then estimate the volume densities of the lines of sight to the Galactic and extragalactic sources, respectively, 
by dividing the measured N$_H^X$ (corrected by the $f_{HI}$ ionization fraction, and, for the Galactic sightlines, for which two absorbers are modeled, weighted by 
the HI ionization fractions of the two components) by $d_{disk}$ (disk WIMM) and $d_{tot}$ (halo WIMM). These density estimates are also reported in Table 9 
for our three Galactic and three extragalactic lines of sight, and their averages are $<n_{H}>_{WIMM}^{Disk} = 0.13 \pm 0.03$ cm$^{-3}$ and $<n_{H}>_{WIMM}^{Halo} = 
0.023 \pm 0.009$ cm$^{-3}$. 
We then conclude that the WIMM absorbing the extragalactic lines of sight is a tenuous and less metal-polluted extension, to large distances above and below 
the Galaxy's disk, and likely to the whole Galaxy's halo, of the WIMM confined in the disk and absorbing the Galactic lines of sight. 

By multiplying the average disk and halo WIMM densities by the disk and halo volumes (i.e. assuming homogenity and isotropy), within our simplified geometry, 
we can derive estimates of the masses of these components. 
We get: M$_{disk}^{WIMM} \simeq 7 \times 10^8$ M$_{\odot}$ and M$_{halo}^{WIMM} \simeq 7.5 \times 10^9$ M$_{\odot}$, and thus 
a total WIMM mass of M$^{WIMM} \simeq 8.2 \times 10^9$ M$_{\odot}$. 

\subsubsection{The Cold Neutral Metal Medium (CNMM) in the Galaxy Disk}
The remaining $\simeq 2/3$ of the OI absorption seen against the Galactic lines of sight is due to a cold gaseous component (T$_{CNMM} \simeq 30-100$ K: 
hereinafter Cold Neutral Metal Mediun, or CNMM), confined in the disk of the Galaxy and characterized by a relatively high metallicity ($<Z>_{CNMM} =( 1.6 \pm 
0.4) Z_{\odot}$ (consistent with the average metallicity of the disk WIMM) and large columns of OI ($<$N$_{OI}>^{CNMM} = (2 \pm 1) \times 10^{18}$ cm$^{-2}$) 
that, however, produce little or no saturation in their K$\alpha$ line, because of the large Doppler parameters ($<b>^{CNMM} = 250 \pm 20$ km s$^{-1}$). 
These large Doppler parameters result from the need to produce large OI K$\alpha$ absorption EWs ($\simeq 2/3$ of this line must be filled in by CNMM 
absorption) without strongly saturating this transition, i.e. overpredicting the CNMM OI column. CNMM OI columns are limited by the 
condition that we impose on the sum of the two {\em galabs} equivalent H column densities (to be linked to the {\em tbabs} equivalent H column of neutral 
H set by the {\em tbabs} best-fitting N$_H^X$) and by the strong anti-correlation between OI columns and Doppler parameters, which would otherwise 
lead to a fast divergence of the O metallicity towards unrealistically high super-solar values. 
Such large $b^{CNMM}$ values cannot be due to intrinsic thermal turbulence within the gas clouds (which are instead cold and should produce oxygen thermal 
broadening not larger than $b_{therm } = 10$ km s$^{-1}$), and are therefore likely casued by large line-of-sight intra-clouds velocity dispersion (we note that 
$<b>^{CNMM} = 250 \pm 20$ km s$^{-1}$ is of the order of the galaxy rotational velocities). 
We thus speculate that the CNMM metal absorption is imprinted by a large number of metal-rich neutral clouds of gas, each of which is intrinsically optically thin 
to OI K$\alpha$ opacity. 

The minimum number of such clouds can be estimated by imposing that the optical depth of each CNMM cloud at the center of the OI K$\alpha$ transition 
be $< 1$. In order to be so, the maximum line-of-sight column density of a single cloud can be at most N$_{OI} \simeq 10^{16}$ cm$^{-2}$. 
This gives a minimum number of clouds $\mathcal{M}>$  $(\overline{N_{OI}^{CNMM}} / 10^{16}) \simeq 2 \times 10^{18} / 10^{16} = 200$, over an average pathlength of 
$<d_{disk}> = 4.6$ kpc. 
The maximum diameter of each CNMM cloud is thus $(<d_{disk}>^{Eff} / \mathcal{M}) = [(<d_{disk}> f_{lin}) / \mathcal{M}] \ls  (<d_{disk}> / \mathcal{M}) \simeq 25$ pc, 
where $d_{disk}^{Eff}$ is the effective pathlength covered by the CNMM clouds along the line of sight and $0 < f_{lin} < 1$ is the line of sight filling factor of the clouds. 

Finally, by dividing the average observed $N_H^X$ by the average $d_{disk}^{Eff}$, we can get an estimate of the average CNMM volume density and so derive 
an estimate of the CNMM mass. We get $<n_{H}>_{CNMM} / f_{lin} = 0.13 \pm 0.03 f_{lin}^{-1} \gs  0.1$ cm$^{-3}$, and thus M$_{CNMM} = 8 \times 
10^8 f_{lin}^2$ M$_{\odot} \ls 8 \times 10^8$ M$_{\odot}$. 


\subsection{Comparison with Previous Works}
Our findings for the Galaxy's disk gaseous components are consistent with those of Pinto et al. (2013), who studied the RGS spectra of 9 
low-mass X-ray binaries. 
They found absorption by cold neutral gas with average metallicity $<Z>_{CNMM}^{Pinto+13} = (0.9 \pm 0.1) Z_{\odot}$, lower than our average 
CNMM value $<Z>_{CNMM} = (1.6 \pm 0.4) Z_{\odot}$, but consistent with that at 1.3$\sigma$ confidence level.  
Pinto et al. (2013) also report on higher ionization metal absorption, including the OII absorption considered here, and find 
OII column densities in the disk of our Galaxy of the order of those found here. However, in their work, the authors fit the ionized 
lines with an ion-by-ion model and therefore do not constrain directly the physical state of the mildly ionized absorber (our disk WIMM), 
nor quantify the fraction of neutral absoprtiopn that also produces absorption by mildly ionized species. As a consequence Pinto et al., 
do not provide an estimate for the metallicity of the disk WIMM. 

\noindent
Yao \& Wang (2006) and Yao et al. (2009) measure the O abundance of the cold (CNMM) and warm (WIMM) ISM phases along two particular lines of sight, those of 
4U~1820-303 and Cygnus X-2. For 4~U1820-303 (not in our sample) the authors neatly separate a cold component (OI) from a warm component (OII-OIII), not 
allowing for part of the OI to be shared by CNMM and WIMM. For Cygnus X~2 (in our sample, but among the lowest S/N spectra), conversely, the authors define a 
single warm component that produces all low-ionization (OI-OIII) oxygen (our Model A). When converted to our A$_O = 4.9 \times 10^{-4}$ units, the  
CNMM and disk-WIMM metallicities reported for these two lines of sight, are: $Z_{CNMM}^{4U 1820-303} = 0.5_{-0.3}^{+1.0} Z_{\odot}$ and $Z_{WIMM}^{4U 1820-303} = 
1.9_{-0.8}^{+3.5} Z_{\odot}$ (when contamination by the OII compound is partly taken into account: Yao \& Wang, 2006) and $Z_{CNMM+WIMM}^{Cyg X-2} = 1.2 \pm 0.3 Z_{\odot}$. 
These measurements agree with our average metallicity estimates for the CNMM and the disk WIMM, within their 1$\sigma$ uncertainties.

As for our estimates of the physical and chemical properties of the halo WIMM of our Galaxy, this is the first time that such estimates are attempted 
through the study of metal X-ray absorption against extragalactic targets. Similar studies, however, have been performed in the Far-Ultraviolet (FUV) 
in the circum-galactic medium of external galaxies (e.g. Lehner et al., 2013). These authors find that the metallicity of the mildly ionized medium ($T\ls 10^4$ K) 
medium surrounding galaxies at z$\ls 1$ has a bimodal distribution, with peaks at [X/H]$\simeq -1.6$ and [X/H]$\simeq -0.3$, and propose that these 
bimodality is due to a galaxy orientation effect, with respect to the line of sight passing through the absorber: the low metallicity systems are those tracking 
pristine gas that is accreting onto galaxies along the galaxy disk planes, while the high metallicity absorbers probe gas reprocessed by the galaxies and 
flowing out of them in the direction perpendicular to their disk planes. 
The average value that we find for [O/H] in the halo-WIMM of our Galaxy ($<Z_{Exgal}>^{WIMM} = (0.4 \pm 0.1) Z_{\odot}$), is fully 
consistent with that of the the high metallicity portion of the bimodal distribution reported by Lehner et al. (2013) for the circum-galactic medium of galaxies 
at $z\ls 1$, possibly indicating that this gas has been first chemically reprocessed in the disk of our Galaxy, and then ejected out of the disk and into the galaxy's 
halo by supernova-driven winds and/or past nuclear activity. 

\section{Conclusions}
We reported on the spectral analysis of two samples of Galactic X-ray binaries and AGNs and the search for low-ionization OI-OII absorption 
at $z\simeq 0$ in these spectra. 
For all targets of our two samples we were able to evaluate the presence and strength of the K$\alpha$ transition of OI (not the K$\beta$ because it falls 
inside the OI K edge trough where the vast majority of our spectra have S/N too low to search for such feature) and two transitions of OII (K$\alpha$ and 
K$\beta$). This allowed us to evaluate saturation and so to remove the degeneracy between ion column density and Doppler parameters and to derive the 
physical, chemical and dynamical properties of the absorbers. 

Our main findins are: 

\begin{enumerate} 
\item Three distinct OI-OII bearing gaseous components are found along the Galactic and extra-galactic lines of sight, respectively. 
We dub these components Cold Neutral Metal Medium (CNMM), Disk Warm Ionized Metal Medium (Disk WIMM), and Halo-WIMM. 

\item The Halo-WIMM is responsible for virtually all the OI-OII absorption seen along the lines of sight to our extragalactic sources, while the Disk-WIMM 
is responsible for virtually the whole OII seen along the lines of sight to our Galactic sources but contributes to only about 1/3 of the OI absorption in the 
Galaxy's disk. 
Halo- and Disk-WIMM are relatively warm ($<$T$_{WIMM}> = 3000 \pm 1000$ K), but the Disk-WIMM is denser, more metal polluted and has larger velocity 
dispersion than the Halo-WIMM: $<n_{H}>_{WIMM}^{Disk} = 0.13 \pm 0.03$ cm$^{-3}$, $<Z_{Gal}>^{WIMM} = (0.8 \pm 0.1)$ $Z_{\odot}$ and 
$<b>_{WIMM}^{Disk} = 100 \pm 10$ km s$^{-1}$ versus $<n_{H}>_{WIMM}^{Halo} = 0.023 \pm 0.009$ cm$^{-3}$, $<Z_{Exgal}>^{WIMM} = 0.4 \pm 0.1$ $Z_{\odot}$ 
and  $<b>_{WIMM}^{Halo} = 48 \pm 6$ km s$^{-1}$, respectively. 
We therefore speculate that the Halo-WIMM is a tenuous and less metal-polluted extension, to large distances above and below 
the Galaxy's disk (and likely to the whole Galaxy's halo) of the Disk-WIMM. 

\item The CNMM absorbs all the Galactic lines of sight but produces little imprints along the high latitude extragalactic lines of sight, and is therefore 
confined in the disk of our Galaxy. 
This medium is cold (T$_{CNMM} \simeq 30-100$ K) and neutral ($<f_{OII}/f_{OI}>_{CNMM} \ls 10^{-4}$), has a relatively high metallicity of $<Z>_{CNMM} = 1.6 
\pm 0.4$ $Z_{\odot}$ (consistent with that of the disk WIMM at 90\% confidence level) and a large average line-of-sight OI coumn density $<$N$_{OI}>^{CNMM} = (2 \pm 1) 
\times 10^{18}$ cm$^{-2}$ combined with an average line-of-sight velocity dispersion of $<b>_{CNMM} = 250 \pm 20$ km s$^{-1}$, of the order of the 
rotational velocities in the Galaxy's disk. 
This let us to the conclusion that the CNMM must be patchy and made up of a large number of clouds ($\gs 40$ kpc$^{-1}$), with an average 
diameter of $\ls 25$ pc each. 

\item By appoximating the disk and halo of of our Galaxy as a geometrically thin cylinder centered on the Galaxy's center,  with radius $R_G = 15$ kpc and 
half-height $h = 0.5$ kpc, and a 15 kpc radius sphere centered in the Galaxy's center, respectively, we were able to estimate the masses of 
the two components: we get: M$_{CNMM} \ls 8 \times 10^8$ and M$_{WIMM} \simeq 8.2 \times 10^9$ M$_{\odot}$ (only 10\% of which in the disk, while the 
remaining 90\% in the halo). 
\end{enumerate}

\section{Acknowledgements}
We thank the anonymous referee for the useful comments that helped improving the paper, and Ehud Behar for providing wavelength, oscillator strengths and 
transition probabilities of inner shell transitions of C, O and Ne. 
FN acknowledges support from INAF-PRIN grant 1.05.01.98.10. 


\end{document}